\documentclass[useAMS,usenatbib,usegraphicx]{mn2e}

\newcommand{\etal}{{\em et al.}}
\newcommand{\atrous}{{\em \`a trous~}}

\title[Intra-group light in compact groups]{Intra-group diffuse light in 
compact groups of galaxies. \\ HCG 79, HCG 88 and HCG 95}
\author[Da Rocha \& Mendes de Oliveira]{C. Da Rocha$^{1,2}$\thanks{Based
on observations obtained at the Canada-France-Hawaii Telescope (CFHT)
which is operated by the National Research Council of Canada, the
Institut National des Science de l'Univers of the Centre National
de la Recherche Scientifique of France, and the University of
Hawaii.}\thanks{E-mail: rocha@astro.physik.uni-goettingen.de} and
C. Mendes de Oliveira$^{3}$\thanks {E-mail: oliveira@astro.iag.usp.br}\\
$^{1}$Institut f\"ur Astrophysik, Friedrich-Hund-Platz 1, 37077
  G\"ottingen, Germany\\
$^{2}$Divis\~ao de Astrof\'{\i}sica, Instituto Nacional de Pesquisas
  Espaciais (INPE/MCT) \\ Av. dos Astronautas 1758, 12227--010,
  S\~ao Jos\'e dos Campos -- SP, Brazil\\
$^{3}$Instituto de Astronomia, Geof\'{\i}sica e Ci\^encias
  Atmosf\'ericas, Universidade de S\~ao Paulo, \\ Rua do Mat\~ao 1226,
  Cidade Universit\'aria, 05508--900, S\~ao Paulo -- SP, Brazil}
\begin{document}

\date{}

\pagerange{\pageref{firstpage}--\pageref{lastpage}} \pubyear{2005}

\maketitle

\label{firstpage}

\begin{abstract}
Deep $B$ and $R$ images of three Hickson Compact Groups, HCG 79, HCG 88
and HCG 95, were analyzed using a new wavelet technic to measure possible
intra-group diffuse light present in these systems.  The method used,
OV\_WAV, is a wavelet technic particularly suitable to detect low-surface
brightness extended structures, down to a $S/N = 0.1$ per pixel, which
corresponds to a 5-$\sigma$-detection level in wavelet space. The
three groups studied are in different evolutionary stages, as can be
judged by their very different fractions of the total light contained
in their intra-group halos: $46\pm11$\% for HCG 79 and $11\pm26$\% for
HCG 95, in the $B$ band, and HCG 88 had no component detected down to
a limiting surface brightness of $29.1~B~mag~arcsec^{-2}$.  For HCG 95
the intra-group light is red, similar to the mean colors of the group
galaxies themselves, suggesting that it is formed by an old population
with no significant on-going star formation.  For HCG 79, however, the
intra-group material has significantly bluer color than the mean color
of the group galaxies, suggesting that the diffuse light may, at least
in part, come from stripping of dwarf galaxies which dissolved into the
group potential well.
\end{abstract}


\section{Introduction}

Galaxy interactions can strip matter from galaxies, forming tidal debris
such as tails, shells and bridges. When the interacting galaxies are
fairly isolated, this material tends to fall back onto the parent galaxy.
First the most bound, lowest angular momentum and later the loosely bound,
high angular momentum material is reaccreted \citep{hib95,mih03}. Stars
may form loops and shells while the gas may form rotating warped disks,
as, for example, in the case of Cen A \citep{mih96,naa01,bar02}. The
gas in those features can also form stars, giving rise to structures
like tidal dwarf galaxies -- TDG's \citep*{zwi56,sch87,mir91,bra00}.

Inside a dense structure, such as in a galaxy cluster, the material
is easily stripped from the member galaxies and the tidal features
are quickly dispersed by the combination of the cluster tides and ram
pressure stripping from the hot intracluster medium, being only partially
re-absorbed by the individual systems. The dispersed matter from the
tidal features will, over time, settle onto the cluster potential well
and form a very faint intracluster diffuse light component \citep{mih03}.

The first mention to intracluster light was by \citet{zwi51}, to describe
an extended envelope of low surface brightness intergalactic matter
observed in the Coma cluster. Other studies have detected diffuse
light envelopes in clusters, such as \citet*{dev60,uso91,sch94},
and more recently \citet{gon00,fel04,ada05,zib05} \citep[See][ for a
review]{vil99}. The intracluster light can represent 10 to 50\% of the
total cluster light, but the measurements have large uncertainties.
In addition, low surface brightness arcs \citep{tre98,gre98,cal00},
individual stars \citep*{fer98}, novae \citep*{nei05} and planetary
nebulae \citep*{fel98,arn02} are being observed in clusters and share
the same origin as the diffuse envelope: stripped material from the
cluster galaxies.

The intracluster light offers a direct way to study the dynamical
evolution of structures, enabling the study of past galaxy encounters,
the dark matter content, history of accretion onto the system and tidal
stripping \citep{dre84}. The amount of intracluster light can indicate
the evolutionary stage of a system, working as an evolutionary clock,
being also sensitive to dark matter distribution, since the shape
of the dark matter halo affects the amount of stripped matter. The
presence of substructures such as low surface brightness arcs gives us
information about the orbits of the galaxies involved in the interaction
episodes. Other unanswered questions such as the origin of the Ultra
Compact Dwarf Galaxies -- UCDs \citep{dri03} and the formation mechanism
for S0 galaxies \citep*{qui00} may also be related to the intracluster
light.

Simulations and theoretical work
\citep*{gal72,mer83,ric83,moo98,kor01,dub03,mih03,gne03,nap03,mur04,wil04,som05}
have greatly increased our knowledge on the formation of the intracluster
light in clusters and groups. It is still unknown, however, how the
specific properties of the intracluster light component correlate with
cluster properties, given the large uncertainties involved and the
few numbers of studies available in this area. It is fair to say that
we are still in the process of forming a global picture, gathering all
information on cluster/group dynamical stages yielded by measurements of
intracluster light. One important addition to the literature data on this
area is the determination of the levels of background light in selected
compact groups of galaxies. This is one of the main contributions of
this paper.

Compact groups of galaxies are very dense environments (projected
densities between $300$ and $10^8~h^2~gal~Mpc^{-2}$), with low velocity
dispersions \citep[$\sim 200~km~s^{-1}$,][]{hic92}, where interactions
should be frequent and the stripping of material from the galaxies should
be an efficient process.  A simple visual inspection of compact groups of
galaxies from Hickson's catalogue \citep[Hickson Compact Group catalogue
-- HCG,][]{hic82} suggests that several of them may contain a diffuse
background light envelope.

Previous studies of compact group galaxies using photographic plates
were unsuccessful in detecting the intra-group light component
(IGL).  \citet{ros79} did not detect any diffuse component and
\citet*{pil95a} have detected an IGL component in only one of the
seven groups studied. A diffuse component was detected by \citet{sul83}
and \citet*{mol98} and more recent studies using CCDs, \citet{nis00}
and \citet{whi03}, have identified IGL in HCG 79 and HCG 90, which
corresponded to 13\% and 45\% of the total light of the group in the $B$
band, respectively.  These previous studies measured the background light
by modeling the galaxies with elliptical isophotes, removing them from
the images and measuring what was left behind. This poses difficulties
first of all because the interacting galaxies are often not well fit by
elliptical isophotes and second because the IGL, being a very low surface
brightness component, usually at a level of less than 1\% of the night
sky brightness, and being typically distributed over a large area of the
group, can be easily contaminated by the light from the group galaxies.

In this paper we describe a new method to isolate the contribution
from the IGL in compact groups, the OV\_WAV \citep*{epi05}, which uses
the \atrous wavelet transform to perform this separation. This method
is independent of galaxy and star modeling and sky level subtraction.
Our method is similar to the one recently applied by \citet{ada05} to
study intracluster light in the Coma cluster. We have applied the method
to images of three compact group galaxies: HCG 79, HCG 88 and HCG 95.
In \S 2 we present the observational data. Section 3 presents the data
reduction with a brief description of the OV\_WAV modeling and the tests
applied to the data. In \S 4 the analysis of the data and the results
are presented. A discussion and summary are presented in section 5.
Throughout this work we use $H_0 = 70~km~s^{-1}~Mpc^{-1}$, $\Omega_M=0.3$,
$\Omega_{\Lambda}=0.7$, when necessary.

\section{Sample and Observational Data}

\subsection{The Sample}

We have selected three groups from Hickson's catalogue \citep{hic82}
which seem to be in different evolutionary stages: 1) HCG 79, which
contains many signs of galaxy interaction and an IGL component which
is readily noticeable by eye inspection, 2) HCG 88, with very few signs
of interaction, despite its very small 2D velocity dispersion of 31 km
s$^{-1}$, 3) HCG 95, which presents clear signs of recent interaction
such as tails and bridges.

\subsubsection{HCG 79}

HCG 79, also known as ``Seyfert Sextet'', was originally identified as a
sextet of galaxies \citep{sey48} and later catalogued by \citet{hic82}
as a quintet. The sixth object is in fact a luminous tidal debris. A
decade later, this group was shown to be a quartet with a mean recession
velocity of $4371~km~s^{-1}$ and a discordant recession velocity galaxy
at $19809~km~s^{-1}$ \citep{hic92}. The velocity dispersion of the group
is $121~km~s^{-1}$.  At a distance of $62.4~Mpc$ ($(m-M)_V=34.0\pm0.6$),
the mean separation of the galaxies is only $7.7~kpc$. It is therefore
the most compact group in Hickson's list \citep{hic82}.

The four galaxies present signs of morphological distortions.  There are
signs of a bar (HCG 79B), tidal tails (HCG 79B and HCG 79D), a dust
lane (HCG 79A) and emission in radio, infrared, [N{\sc ii}] and ${\rm
H{\alpha}}$ \citep[HCG 79A and HCG 79B,][]{men94}, disturbed rotation
curves \citep[HCG 79A and HCG 79B,][]{bon99} and nuclear activity (HCG
79A, HCG 79B and HCG 79D).  Signs of the nuclear activity were detected
and classified by \citet{shi00} using [N{\sc ii}], [S{\sc ii}] and
${\rm H{\alpha}}$ ratios.  The group presents a prominent IGL envelope,
possible to be identified by eye inspection and irregular envelopes
of H{\sc i} \citep*{wil91} and X-rays \citep*{pil95b}. These suggest
that recent or on-going interaction has taken place within this system.
The group data are summarized in table~\ref{tabdata}.

\subsubsection{HCG 88}

This system is composed by a quartet of late-type galaxies aligned in a
filamentary structure, with a mean recession velocity of $6040~km~s^{-1}$
\citep{hic92} and a velocity dispersion of only $31~km~s^{-1}$. The
mean separation of the galaxies is $106~kpc$, at a distance of $86.3~Mpc$
($(m-M)_V=34.7\pm0.7$).

The group members present some morphological distortions, such as low
luminosity extensions (HCG 88B). They are detected in ${\rm H{\alpha}}$,
infrared and radio emission \citep{men94} but none of the galaxies
have highly perturbed rotation curves \citep{pla03}. About 90\% of
the detected H{\sc i} content of this group is still associated with
the disks of the galaxies \citep{ver01}.  \citet{dec97} identified two
other (optically fainter) galaxies at the same redshift of the group.
A summary of the group data are presented in table~\ref{tabdata}.

\subsubsection{HCG 95}

This system was catalogued by \citet{hic82} as a quartet of galaxies with
a mean recession velocity of $11859~km~s^{-1}$, a velocity dispersion
of $356~km~s^{-1}$ and a mean separation of $48~kpc$, at a distance of
$169.4~Mpc$ ($(m-M)_V=36.1\pm0.7$).

By simple eye-inspection of a broad-band image of the group one promptly
notices very obvious morphological peculiarities: two tidal tails in the
HCG 95A and HCG 95C region, which seems to connect A and C, and a double
nucleus in HCG 95C. HCG 95C is most probably an on-going merger of two
disk galaxies, also in interaction with HCG 95A (an elliptical galaxy).
\citet{men94} noted radio and infrared emission as well as morphological
distortions in HCG 95B.

Reconstructed IRAS frames show infrared emission in this group centered
on HCG 95C and enclosing the whole group \citep{all96}. An upper limit
to the X-ray detection was given by \citet{pon96} using ROSAT data.
H{\sc i} was also detected in this group \citep{huc00,ver01}.

\citet{vil98} detected ${\rm H{\alpha}}$ emission in the nuclei of HCG
95A and HCG 95C and in one of the large tidal tails (the eastern tail).
Such tail seems to contain a few tidal dwarf galaxies candidates
\citep{igl01}. ${\rm H{\alpha}}$ emission was also detected along the
HCG 95D edge-on disk. HCG 95B (CGCG 406--067) had ${\rm H{\alpha}}$
emission detected along its disk \citep{igl98}, but at a radial velocity
of $8000~km~s^{-1}$, instead of at the group's velocity. This indicates
that this non-member galaxy is in the foreground of the group.  Besides
the discrepant radial velocity of HCG 95B published by \citet{hic92}
($11637~km~s^{-1}$), other similarly erroneous values were published
at the RC3 \citep[$11486~km~s^{-1}$,][]{dev91} and \citet{fou92}
($11562~km~s^{-1}$). However, a spectrum of this galaxy obtained
with the 1.52 meter telescope at ESO on November 2002 confirmed the
result of \citet{igl98}, that its radial velocity is indeed $\sim
8000~km~s^{-1}$ and not around $11500~km~s^{-1}$. According to those
authors, the morphological distortions presented by this galaxy would
have been caused by a nearby dwarf galaxy. The group properties are also
summarized in table~\ref{tabdata}.

\begin{table}

\centering
\caption{General properties of the observed groups.
\label{tabdata}}

\begin{tabular}{llll}

\hline
           & HCG 79               & HCG 88               & HCG 95 \\
\hline

RA         & ${\rm 15^h59^m11^s9}$&${\rm 20^h52^m22^s8}$&${\rm 23^h19^m31^s8}$\\
DEC        & ${\rm +20^o45'31''}$ &${\rm -05^o45'28''}$ &${\rm +09^o29'31''}$ \\
$V_{Rad}$  & 4371 ($km~s^{-1}$)   & 6040 ($km~s^{-1}$)  & 11859 ($km~s^{-1}$) \\
Distance   & 62.4 ($Mpc$)         & 86.3 ($Mpc$)        & 169.4 ($Mpc$)       \\
Vel. Disp. & 121 ($km~s^{-1})$    &  31 ($km~s^{-1}$)   & 356 ($km~s^{-1}$)   \\
Mean. Sep. & 7.7 ($kpc$)          & 106 ($kpc$)         & 48 ($kpc$)          \\
Num. Gal.  & 4                    & 4                   & 3                   \\

\hline

\end{tabular}
\end{table}

\subsection{Observational data}

$B$ and $R$ deep images were obtained at the CFHT (Canada--France--Hawaii
Telescope). The filters used were B and R from the Mould system,
but the resulting magnitudes were transformed into the standard
Johnson-Morgan-Cousin B and R magnitudes using a set of standard stars
obtained in the same nights, as explained below.

Images of HCG 79 were obtained using the SIS (Sub--arcsecond Imaging
Spectrograph), with exposure times of 2700 seconds ($5\times540$) in the
$B$ band and 1800 seconds ($3\times600$) in the $R$ band. The average
seeing values were $0\farcs83$ and $0\farcs64$ for the $B$ and $R$ images,
respectively, and the field size $2\farcm4\times3\farcm1$ with a pixel
size of 0.173 arcsecond (the detector was binned 2 x 2). The photometric
zero points were obtained using standard stars from \citet{lan92} and
from the Galactic globular clusters M92 and NGC 7006 \citep{chr85}.

Images for HCG 88 and HCG 95 were obtained using the MOS (Multi--Object
Spectrograph), with a pixel size of 0.314 arcsecond and a field size of
$9\farcm4\times9\farcm1$.  The exposure times for the HCG 88 images were
$4\times900$ and $8\times600$ for the $B$ and $R$ images respectively,
with a similar mean seeing in both filters of $0\farcs95$. The HCG 95
images had exposure times of $3\times800$ and $10\times300$ for the $B$
and $R$ images respectively, with a similar mean seeing in both filters
of $0\farcs96$.  The photometric zero points were, also in this case,
calibrated using the Galactic globular clusters M92 and NGC 7006
\citep{chr85}.

The characteristics of the images are summarized in table~\ref{tabima}.

\begin{table*}

\centering
\caption{Observational data.
\label{tabima}}

\begin{tabular}{cclrlccl}

\hline
Group & Band   & Inst. & Date & 
Exp.  & Seeing & FOV   & Pixel \\

      &        &       &      & 
(Sec.)&        &       & 
(${\stackrel{''}{}}/pixel$) \\
\hline

HCG 79 & B & SIS/CFHT & Aug, 93 & 2700 ($5\times540$)  & $0\farcs83$ &
  $2\farcm4\times3\farcm1$ & 0.173 \\
       & R & SIS/CFHT & Aug, 93 & 1800 ($3\times600$)  & $0\farcs64$ &
  $2\farcm4\times3\farcm1$ & 0.173 \\
HCG 88 & B & MOS/CFHT & Jul, 94 & 3600 ($4\times900$)  & $0\farcs94$ & 
  $2\farcm4\times3\farcm1$ & 0.314 \\
       & R & MOS/CFHT & Jul, 94 & 4800 ($8\times600$)  & $0\farcs96$ &
  $2\farcm4\times3\farcm1$ & 0.314 \\
HCG 95 & B & MOS/CFHT & Jul, 94 & 2400 ($3\times800$)  & $0\farcs93$ & 
  $2\farcm4\times3\farcm1$ & 0.314 \\
       & R & MOS/CFHT & Jul, 94 & 3000 ($10\times300$) & $0\farcs98$ &
  $2\farcm4\times3\farcm1$ & 0.314 \\

\hline

\end{tabular}
\end{table*}

\section{Data Reduction}

The IGL is a very low surface brightness component, usually at a level
of less than 1\% of the night sky brightness, but typically distributed
over a large area of the group.  Many effects can appear as contaminating
features to the signal one wants to measure.  First those caused by
technical problems at the time one takes the data, such as instrumental
scattering, {\it i.e.}, spurious light inside the telescope, flatfielding,
which must be better than 1\%, ghosts or fringing, vignetting and CCD
bleeding. Other problems that make it difficult to measure the IGL
component: the background light is dimmed by the light from the group
galaxies and bright stars, which should be modeled and removed from the
image; measuring the sky level accurately; contamination from faint
galaxies beyond the detection limit; galaxy absorption and galactic
cirrus.

Assuming one has the technical problems under control, the two main
problems of detecting background light are the sky level subtraction and
modeling of the bright galaxies and stars superimposed onto the group
image. To overcome those problems we have applied a wavelet based technic,
the OV\_WAV package \citep{epi05}, in which the necessary information
to treat those effects is internally determined by the package. The
description of the technic is given below.

A similar technic was recently employed by \citet{ada05} to study
intracluster light in the Coma cluster. This work uses the same wavelet
transform, the {\em \`a trous}, and the same object reconstruction
algorithm, but in our case we have chosen to reconstruct each of the
detected objects individually. With this individual object reconstruction,
we are able to select only the structures which compose the IGL.

The basic image reduction procedures (bias correction, flatfielding and
image combining) were done using the IRAF\footnotemark~package (Image
Reduction and Analysis Facility). The photometric calibrations with
standard stars were performed using {\tt DIGIPHOT.PHOTCAL} under IRAF.
The final combined images were then processed with the OV\_WAV package,
described below, running under IDL (Interactive Data Language - Developed
by {\it Research Systems Inc.} -- RSI).

\footnotetext{IRAF is distributed by the National Optical Astronomy
Observatories, which is operated by the Association of Universities
for Research in Astronomy, Inc., under cooperative agreement with the
National Science Foundation.}

\subsection{The OV\_WAV Package}

The OV\_WAV is a multiscale package for astronomical data reduction,
based on the \atrous wavelet transform \citep{epi05}. Astronomical
images are usually hierarchically organized, {\it i.e.}, smaller
structures are located inside larger structures. Thus, stars may be
projected onto a galaxy, which in turn may be projected onto a large
diffuse envelope (as in our case), which is then projected onto the
sky brightness level. Multiscale analysis is very well suited for this
kind of study, since it can separate the structures found in an image by
their characteristic sizes.  The \atrous wavelet transform is a redundant
discrete transform, appropriate for digital astronomical images, since
it conserves flux and avoids the errors introduced by the discretization
process of a continuous function.

The general procedure is an improved version of the procedure described
by \citet*{sta98}, which consists in deconvolving the signal into wavelet
coefficients (since images are a 2D signal, each wavelet coefficient
corresponds to a plane), identifying the objects representations in the
wavelet space, defining the objects with a ``multiscale vision model''
\citep{bij95} and reconstructing the detected objects.

\begin{figure*}
\centering
\includegraphics[width=5.0cm]{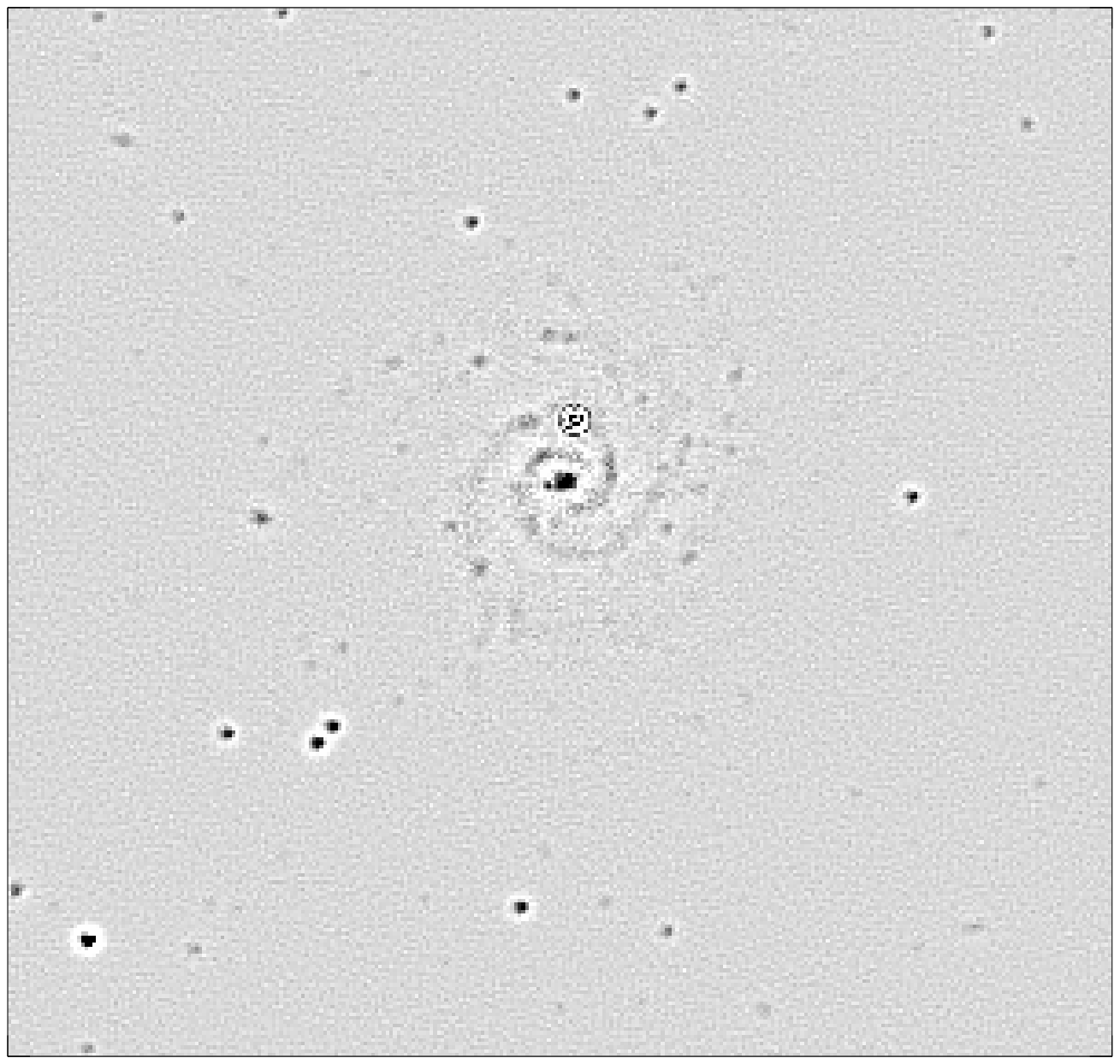}
\includegraphics[width=5.0cm]{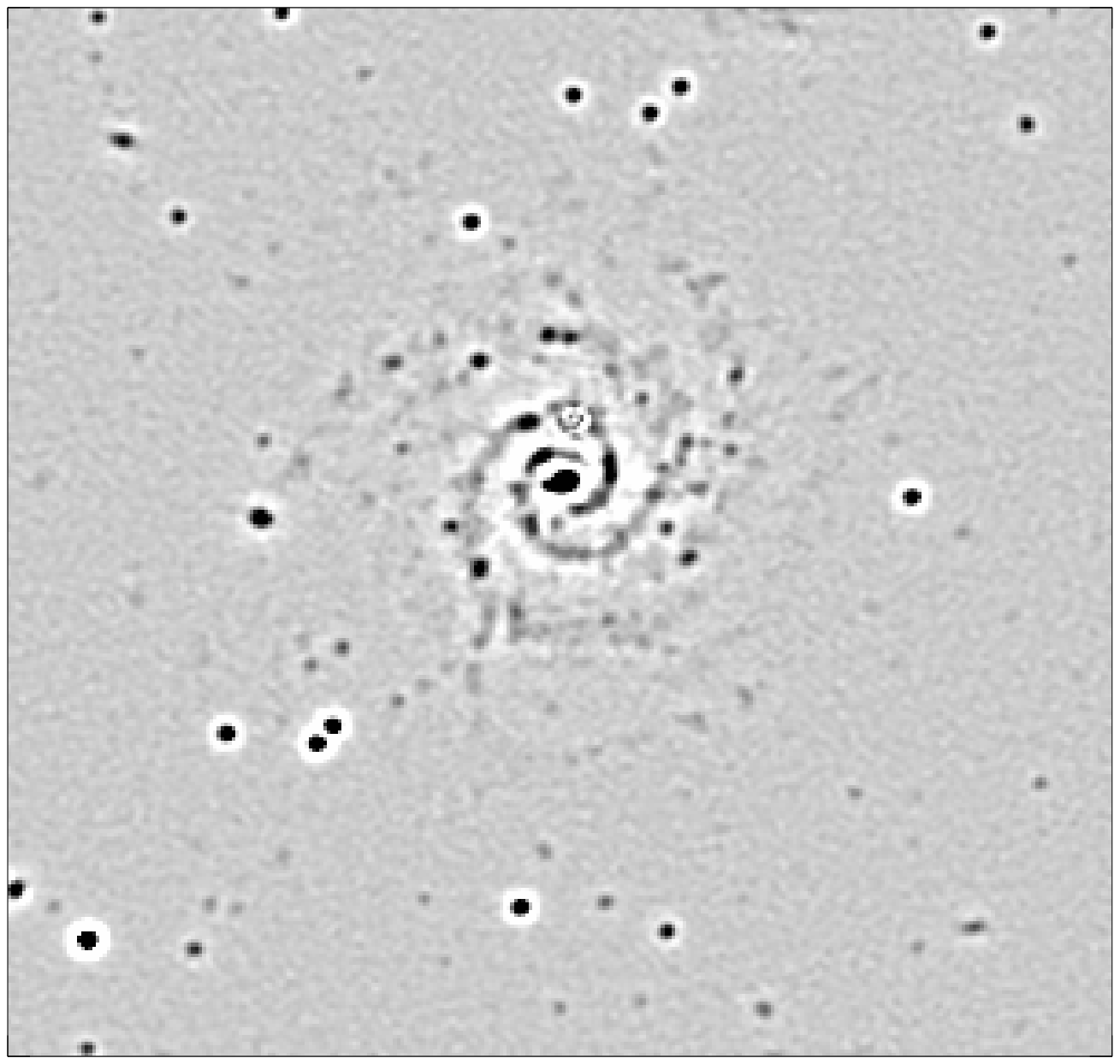}
\includegraphics[width=5.0cm]{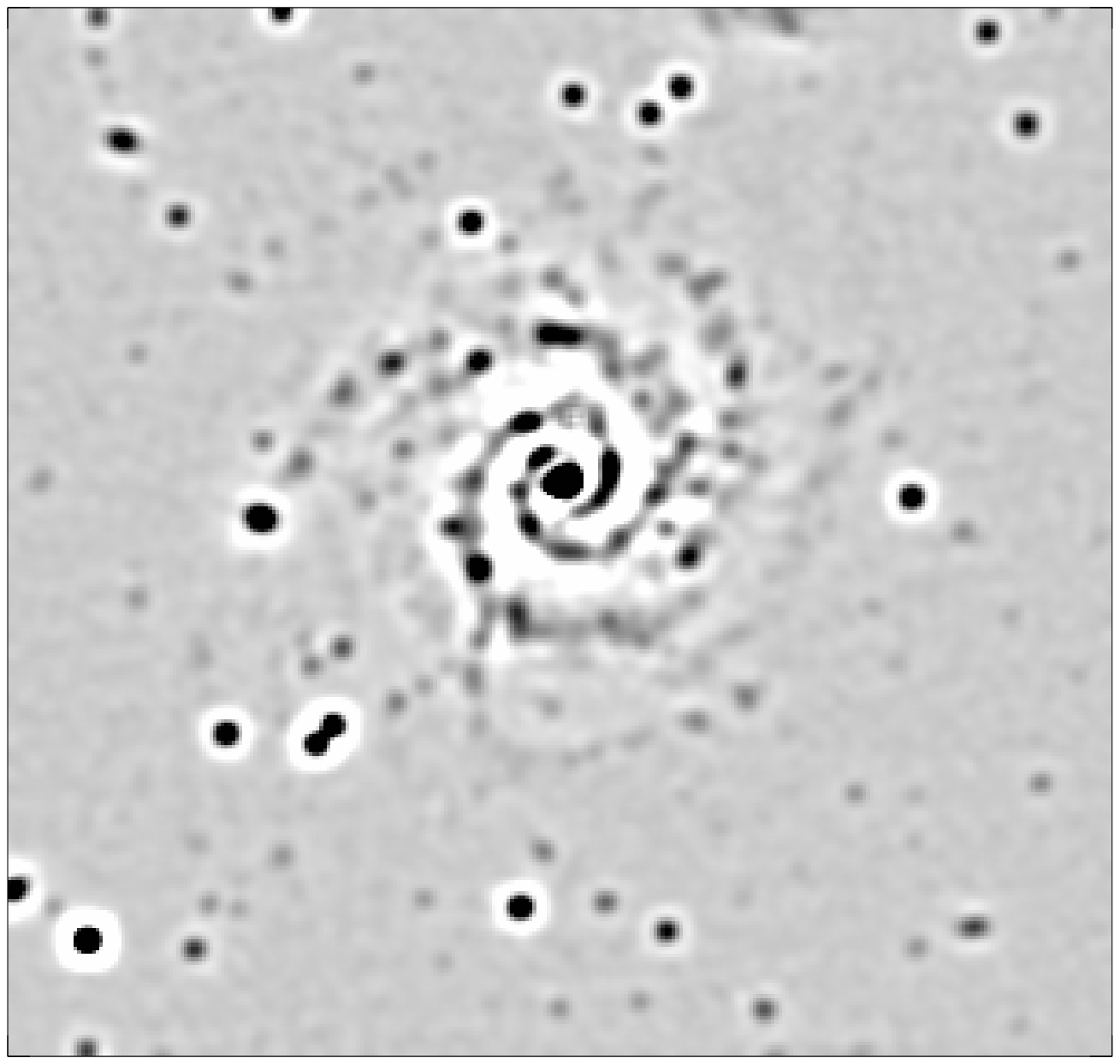}
\includegraphics[width=5.0cm]{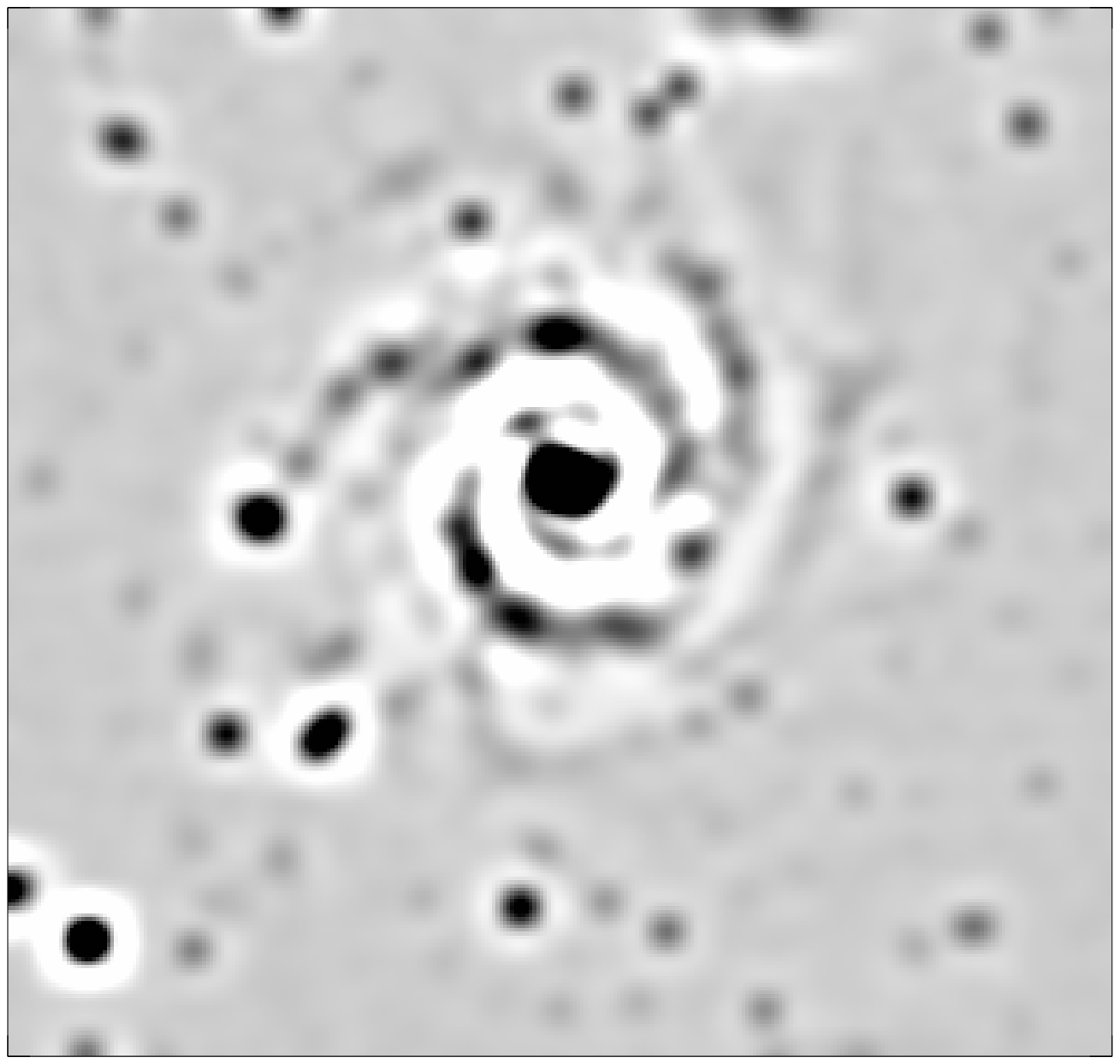}
\includegraphics[width=5.0cm]{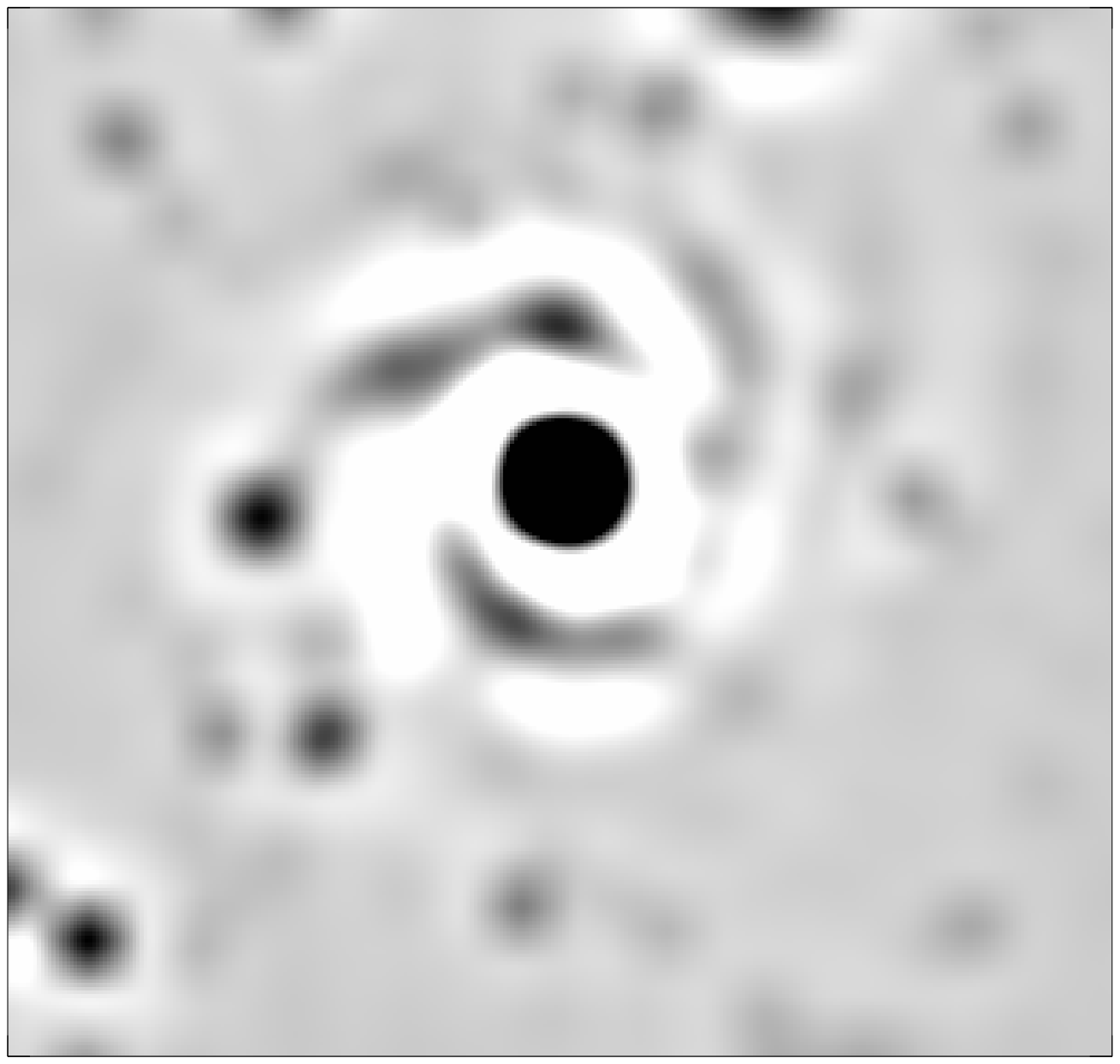}
\includegraphics[width=5.0cm]{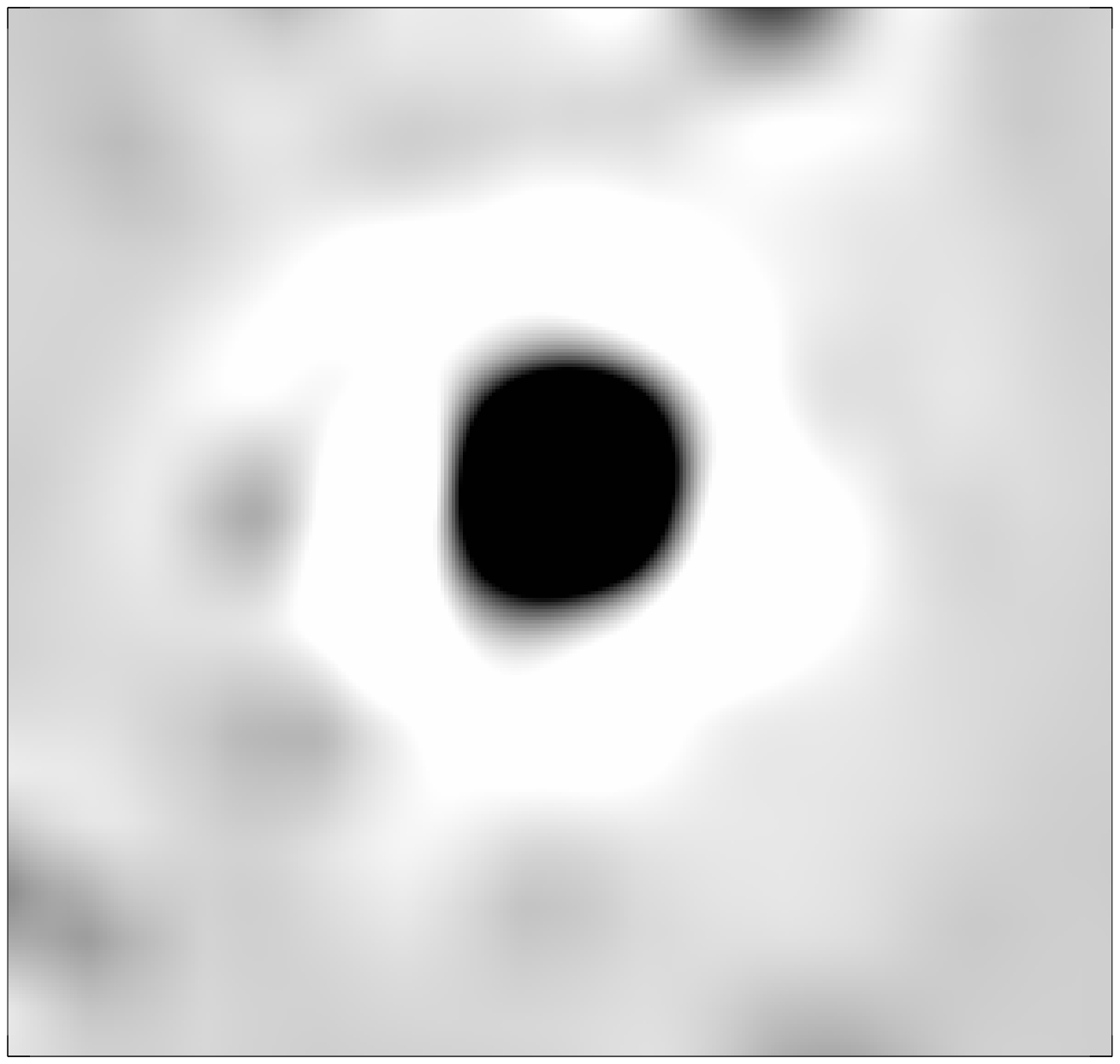}
\includegraphics[width=5.0cm]{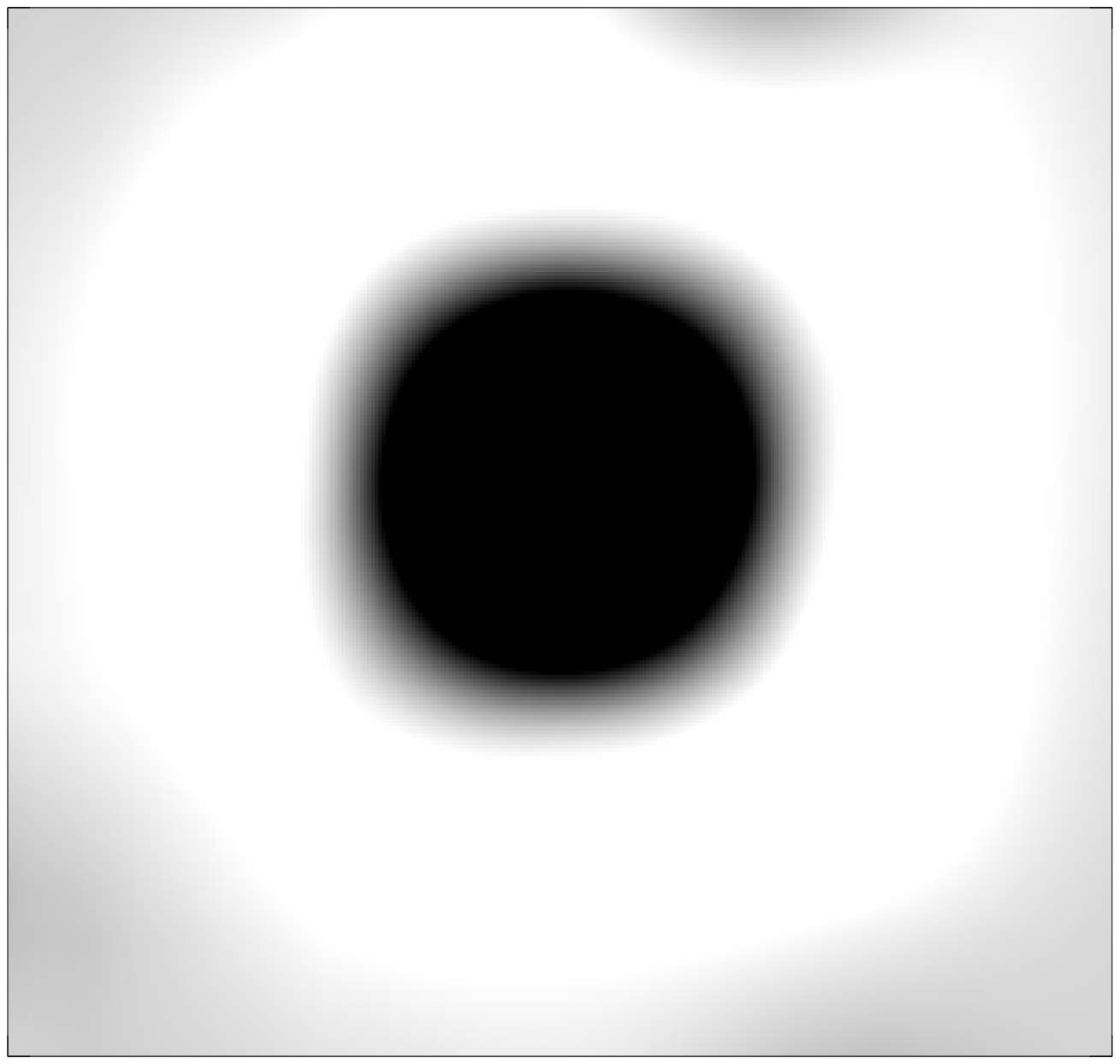}
\includegraphics[width=5.0cm]{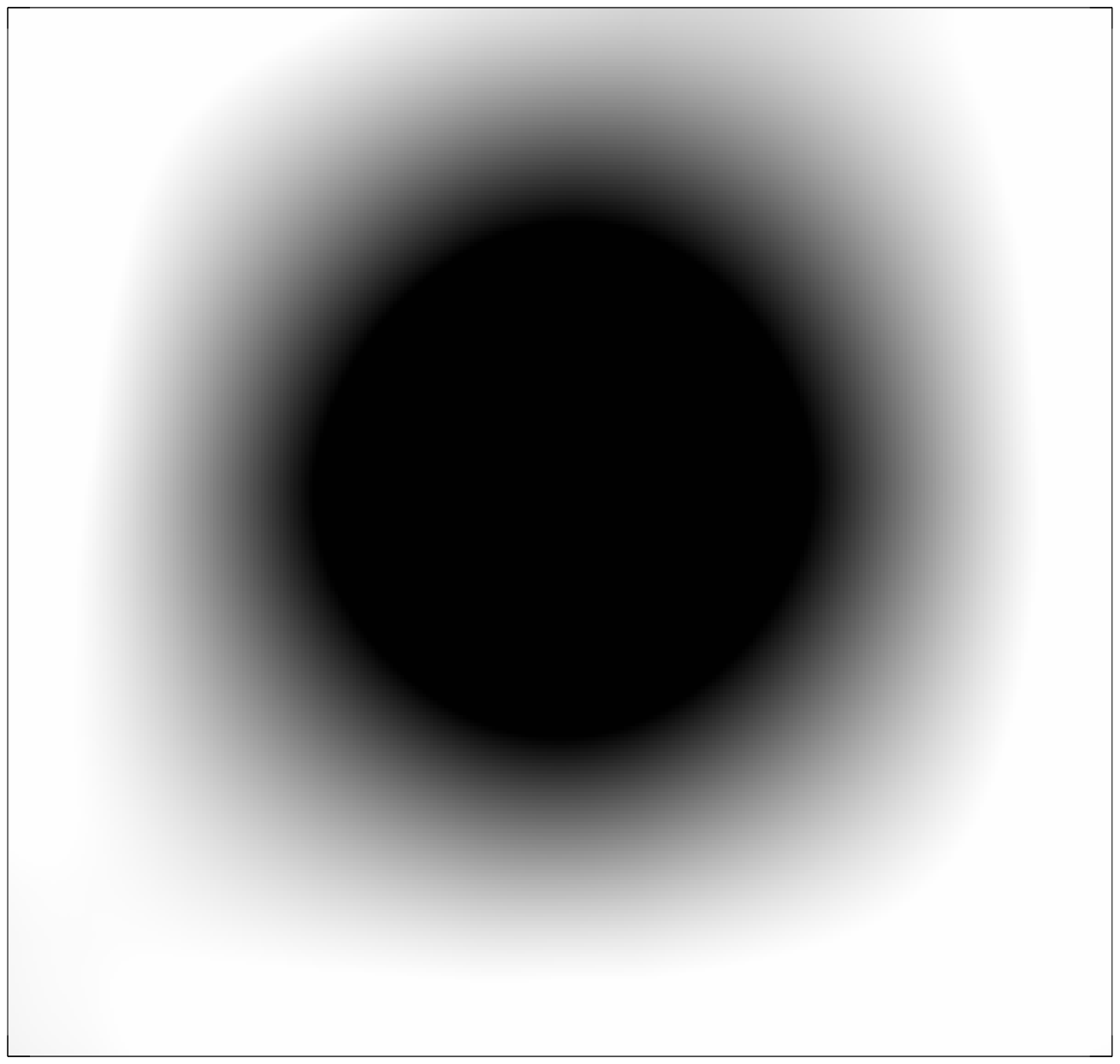}
\includegraphics[width=5.0cm]{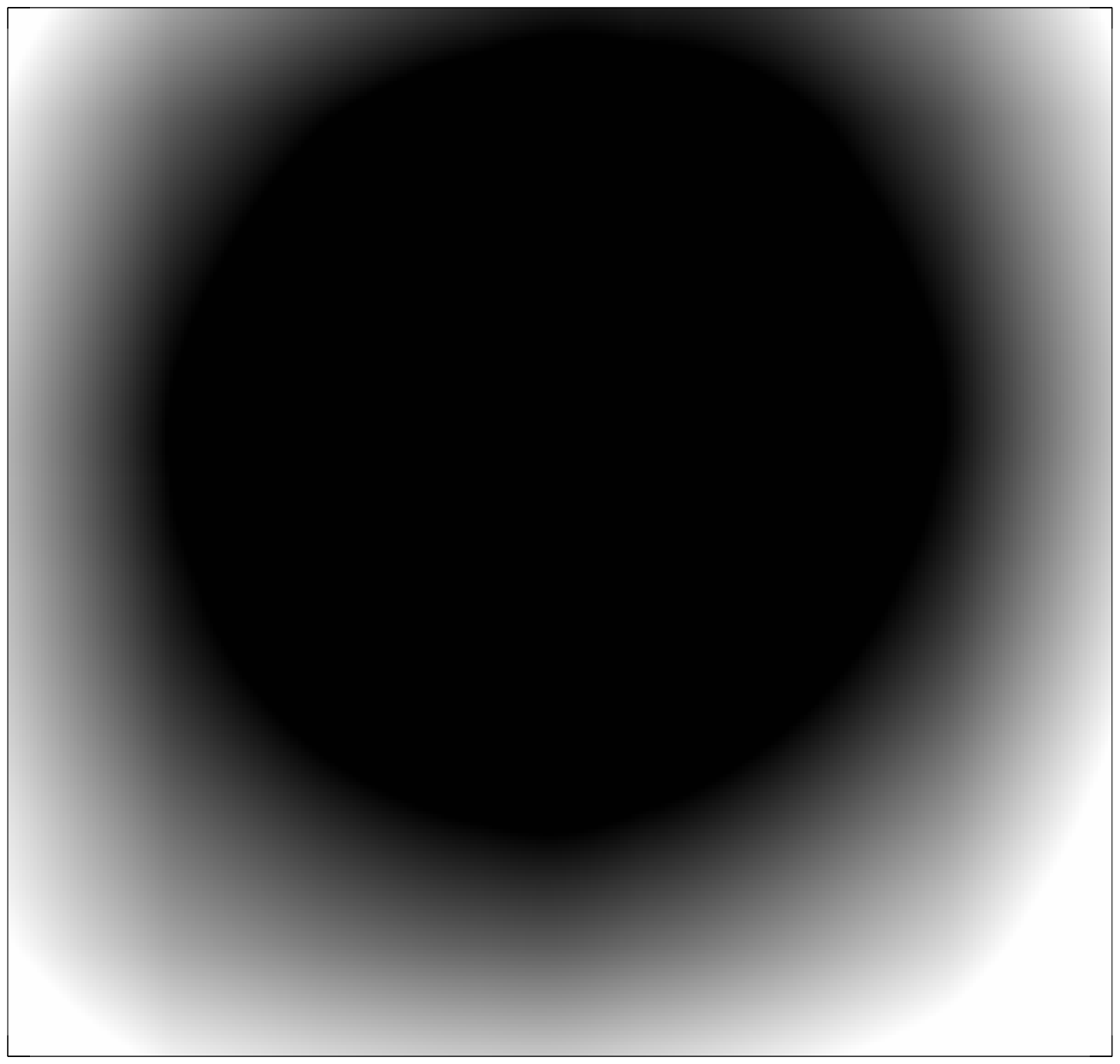}
\caption{Deconvolution of a spiral galaxy (HCG 88C) in 9 wavelet
coefficients as an example of object representation and reduction of the
noise in different coefficients.
\label{figdecon}}
\end{figure*}

A given object is represented in the wavelet space in different subsequent
wavelet coefficients. Those representations are identified in each
coefficient using a threshold, which depends on the representation of the
image noise in each of the coefficients. As each coefficient contains only
structures of a certain size ($2^n$ pixels, where $n$ is the index of the
wavelet coefficient), the representation of the noise in each coefficient
also has the same characteristic size. In the first coefficient ($n=0$)
we only have structures with the size of one pixel, and in this case the
noise is basically the pixel to pixel noise, which is usually high. For
all further coefficients ($n=1, 2, 3, ...$) the representation of the
noise has larger characteristic size ($2$ pixels, $4$ pixel, $8$ pixels,
...) and has a considerably reduced intensity. Extended low surface
brightness objects, which have very low signal to noise ratios ($S/N$) in
normal space, will be identified in the largest coefficients with higher
$S/N$, since they will only be affected by very low intensity noise,
with the characteristic size of the coefficient. By this special relation
between the signal and the noise representations in wavelet space, this
method is very efficient in detecting low surface brightness structures,
which would usually be lost if traditional technics would be employed.
As an example of deconvolution of an object, figure~\ref{figdecon} shows
the representation of a spiral galaxy (HCG 88C) in 9 wavelet coefficients,
where the reduction of the noise with the increase of the characteristic
size of the coefficient can be noticed.

\begin{figure}
\centering
\includegraphics[width=6.0cm]{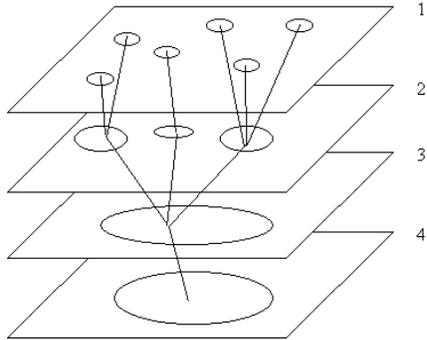}
\caption{Schematic version of a connectivity tree connecting regions
detected in different wavelet coefficients composing an object.
\label{figtree}}
\end{figure}

The object's representations in each coefficient are identified,
defining valid regions, and those regions are interconnected, forming the
object as a whole. The region which contains the pixel with the highest
flux of all other regions of this object will be considered the ``main
region'' of the object and the object will be marked as detected in this
particular coefficient.  This way the detected objects are separated into
characteristic sizes.  With this technic, it is possible to separate
different size objects and structures like point sources, resolved
small sources, larger galaxies, the IGL component and the sky brightness
without any assumption about the shape of the sources or the sky level. A
schematic version of the connectivity tree of regions in different wavelet
coefficients composing the object can be seen in figure~\ref{figtree}.

\begin{figure}
\centering
\includegraphics[width=6.0cm]{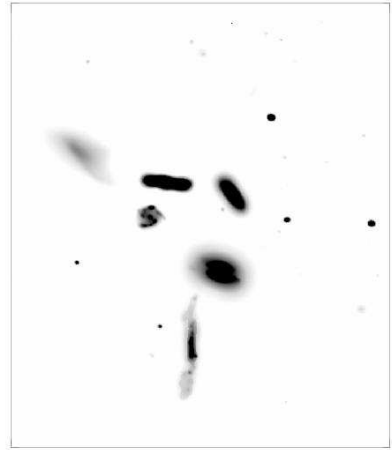}
\includegraphics[width=6.0cm]{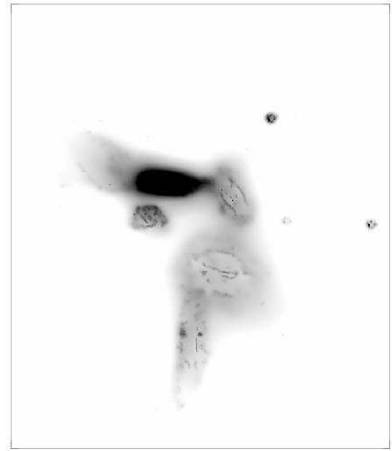}
\includegraphics[width=6.0cm]{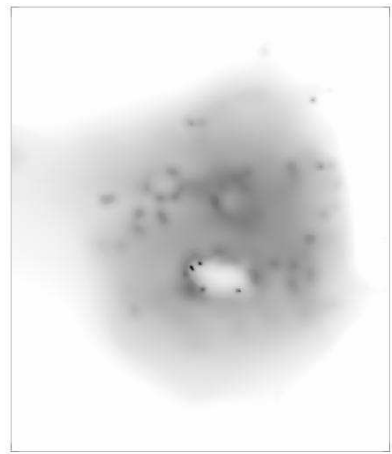}
\caption{Analysis of HCG 79 $B$ image in three iterations. In the
first we can notice the detection of bright structures as the point
sources and the central part of the galaxies. In the second iteration
the wings of the brightest point sources and the halos of the galaxies
were detected. In the third iteration the IGL component and some small
features of the bright sources were detected.
\label{figmultiiter}}
\end{figure}

The reconstruction of an object marked as detected in a certain
coefficient ($z$), will use all the lower coefficients, the main
coefficient and one larger coefficient ($n \le z+1$). This procedure
favors the detection of bright small sources, as point sources and
galaxy central parts, so that the image has to be processed in a
sequence of iterations. In the first iteration we have the detection
of the small bright sources, which are reconstructed and subtracted
from the image. In the second iteration fainter and larger sources are
identified, as for example the outer parts of bright point sources and
galaxies' halos. These are then reconstructed and subtracted from the
image to allow another iteration. In the third iteration the IGL starts
showing up.  An example of multiple iteration image analysis can be seen
in figure~\ref{figmultiiter}. For a complete description of the package
and the method see \citet{epi05}.

\subsection{Package Test and Simulations}

We have performed tests and simulations to define an optimized set of
configuration parameters to detect IGL and the confidence level of our
results with the OV\_WAV.

A B-Spline filter was used for deconvolving the image in wavelet
coefficients, the Anscombe noise transformation from Poissonian to
Gaussian noise was applied and the valid regions were detected at a
5-$\sigma$-detection level.  The convergence reconstruction was reached
when the flux difference between two iterations was smaller than $10^{-3}$
of the last iteration total flux.

\subsubsection{Simulated Images}

Using the {\tt ARTDATA.MKOBJECTS} task, inside IRAF, we have generated
three different configurations of artificial fields to test the setup
of configuration parameters and the confidence level of our results. The
fields are $600\times600~\rm{pixels^2}$ with a Poissonian sky noise.

a) In the first configuration, we have included four galaxies with an
exponential brightness profile, with total magnitudes within a range of
two magnitudes. The galaxies were added within a very small projected
distance from one another, in an attempt to mimic a compact group, with
a maximum distance among the centers of the objects of about 30 pixels
and without any kind of diffuse component. In this case our main goal
was to test if any diffuse-like contamination could be introduced in
our results due to the overlapping of the various galaxy halos.

b) The second configuration was a very extended exponential profile
galaxy, with an effective radius of 50 pixels (the galaxies in
configuration (a) had effective radii or about 3 to 4 pixels). The
extended profile covered most of the image, emulating what we expect for
an IGL component. This configuration had no other galaxies superimposed
onto it, in order to evaluate the IGL detection efficiency without any
contamination effect. This second test was performed with three different
IGL luminosities, with a central ($S/N$) ratio per pixel of $2.50$, $1.25$
and $0.63$ and mean $S/N$ of $0.41$, $0.29$ and $0.21$, respectively,
the latter two being indeed very low signals.  These correspond to cases
A, B and C respectively, in table~\ref{tabsimgloblim}.

c) The third image configuration has the IGL component as simulated in
(b), with the four galaxies of configuration (a) above superposed onto
it. The IGL component was also generated using the three different IGL
luminosities.

\begin{figure}
\centering
\includegraphics[width=6.0cm]{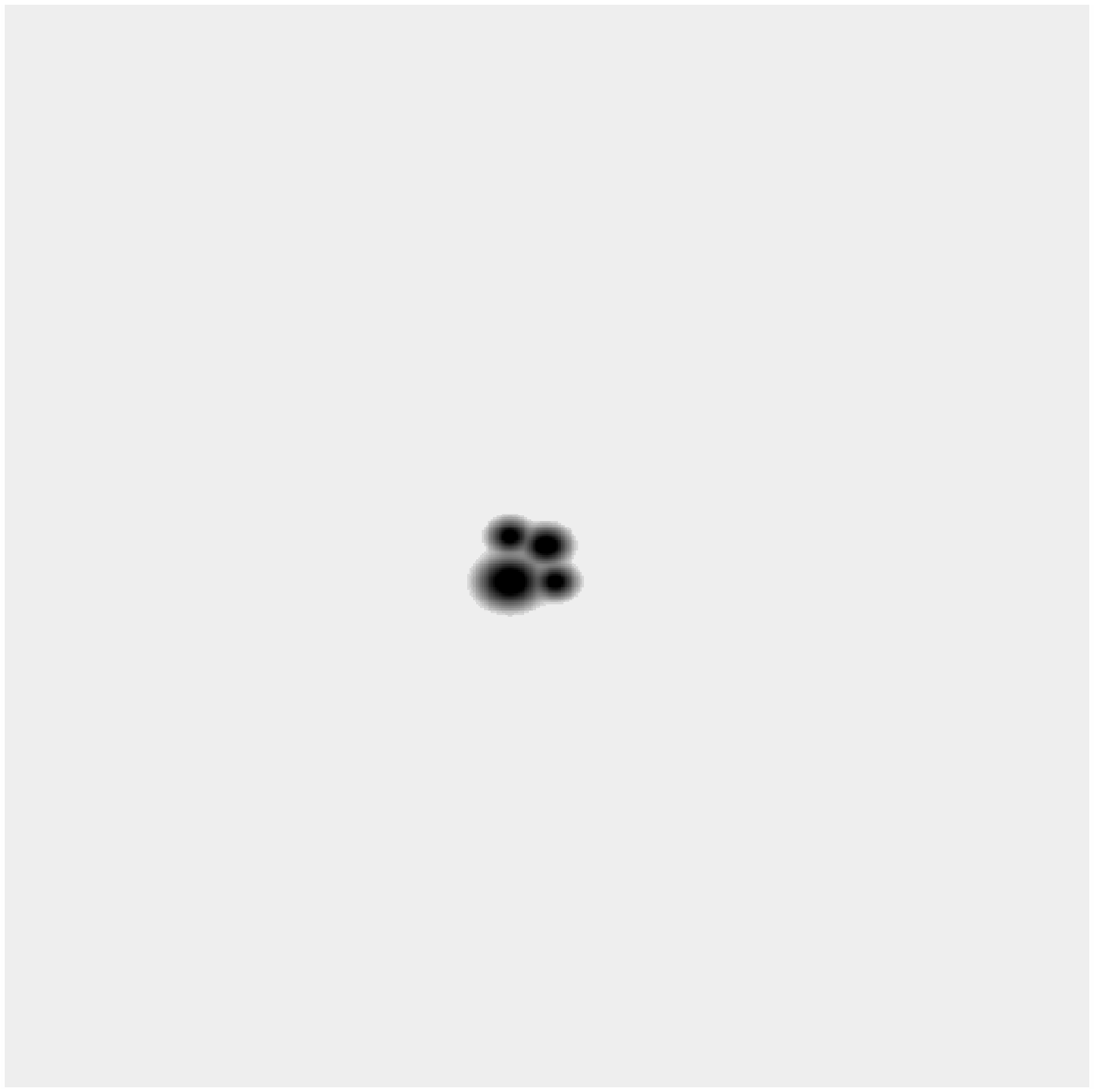}
\includegraphics[width=6.0cm]{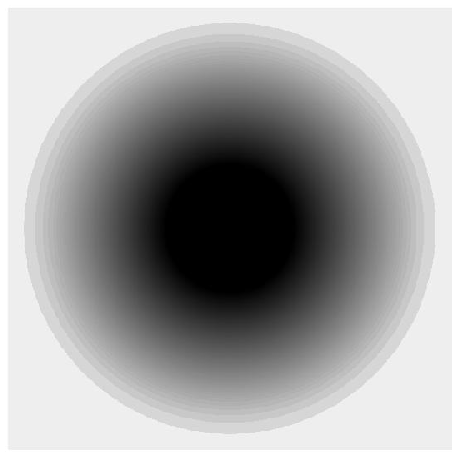}
\includegraphics[width=6.0cm]{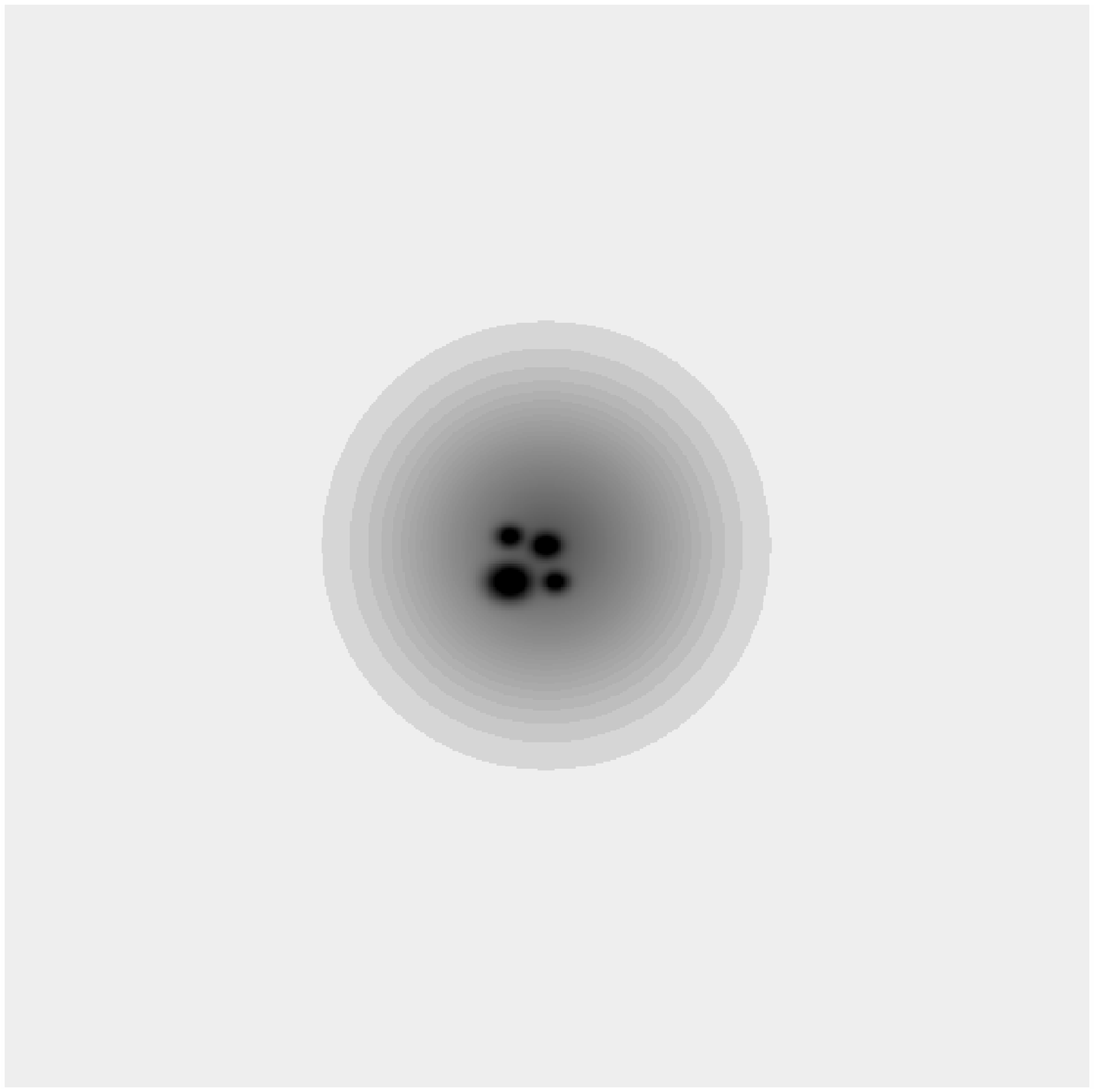}
\caption{The upper panel shows configuration (a), a compact group of four
exponential galaxies without any diffuse component. The middle panel
shows configuration (b), a very extended exponential galaxy, emulating
an intra-group light component. The lower panel shows configuration (c),
the diffuse component with the compact group superposed onto it.  The IGL
component in the lower panel is as extended as in the middle panel,
but the contrast of the display shown here is different.  All figures
are presented here as noiseless versions, to allow visualization of the
faintest regions.
\label{figsimuima}}
\end{figure}

We had, in the end, a set of seven different images to analyze.
The configurations for the three simulated images are shown in
figure~\ref{figsimuima}.

\subsubsection{Analysis Parameters}

The analysis of each image was performed deconvolving it in 10
wavelet coefficients, where the last coefficient corresponds to a
characteristic size of 512 pixels ($2^9$), which is the approximate
size of the image. Two parameters were varied to evaluate the package
performance. One was the object definition criterion for valid regions,
which can be the ``standard'' case, where a minimum of two valid regions
connected on subsequent wavelet coefficients define an object, or the
``restricted'' case, where at least three valid regions connected on
subsequent wavelet coefficients are needed to define an object. The
other parameter is the definition of whether a pixel belongs to a certain
region or not. One possibility is the ``hard thresholding'', where a flag
(0 or 1) is assigned to a pixel, to mark if it belongs to the region or
not. The alternative is the so called ``combined evidence thresholding'',
where a weight (from 0 to 1) is assigned to a given pixel.

Four possible sets of parameters were tested: I -- using restricted
criterion and combined evidence thresholding; II -- using restricted
criterion and hard thresholding ; III -- using the standard criterion
and the combined evidence thresholding; and IV -- using the standard
criterion and the hard thresholding.

The combined evidence thresholding is very sensitive to small size
sky fluctuations that can appear in subsequent wavelet coefficients,
creating spurious small scale object detections.  We have forced the use
of the hard thresholding up to the third coefficient ($2^2$ pixels),
to avoid such a problem. This procedure does not affect the analysis
of the galaxies nor IGL component present in the images, since their
characteristic sizes are much larger than those of the structures detected
in the third-coefficient image.

\subsubsection{Simulation Results}

We first performed a global analysis of the simulated images, evaluating
the differences in the integrated fluxes between the artificial models and
the reconstructed objects and the quality of the reconstruction process.

Using any of the four parameter setups I to IV described above, the
four galaxies of the first image (without diffuse component) can be
perfectly separated. Two possible contaminating effects were identified
using setups I and II (with the restricted criterion). At the fourth
iteration, a ``false'' common halo formed by the overlapping galaxies was
identified as an object, which despite being centered on the galaxies,
could be mistakenly identified as part of the IGL component. A second
larger ``false'' common halo can also be noticed in the reconstructed
objects of a further iteration. With setups III and IV (those without
the restricted criterion) a single ``false'' common halo was identified.
This halo contains almost the same structure identified with setups I
and II as two different objects, but more extended. The more extended
single ``false'' halos identified in setups III and IV, indicate
that those objects were detected at a larger wavelet coefficient,
which would be more difficult to separate from the IGL component in a
real case.  The identification of these different contamination sources
show the superior capability of the restricted criterion in separating
embedded structures offering a more reliable object reconstruction.
The contamination by those artifacts is about 2\% in all cases.

\begin{figure}
\centering
\includegraphics[width=6.0cm]{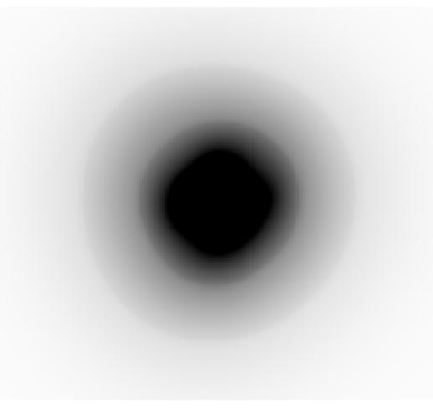}
\includegraphics[width=6.0cm]{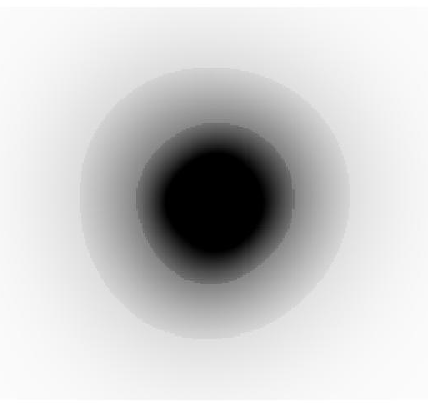}
\includegraphics[width=6.5cm]{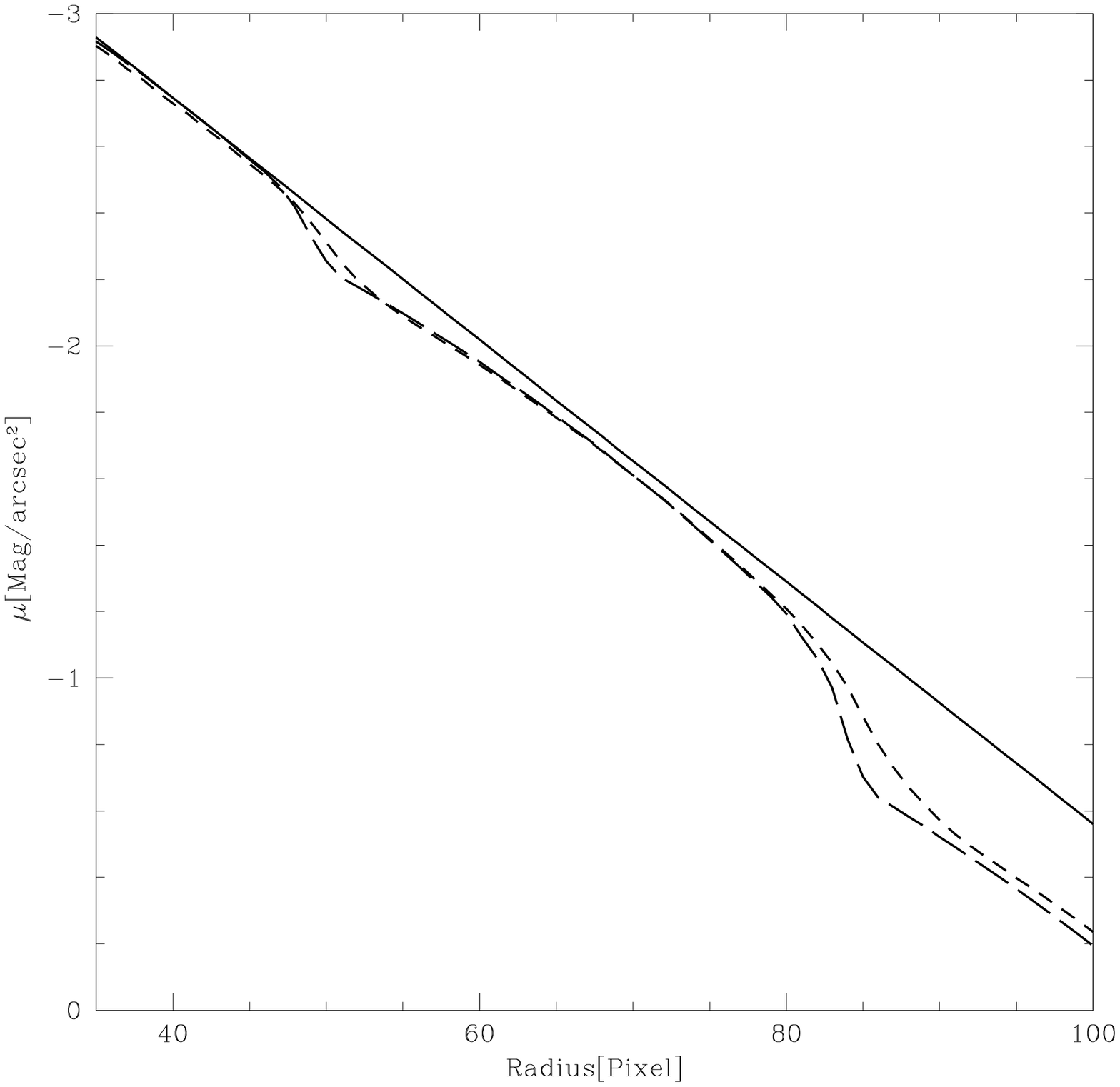}
\caption{The upper panel shows the reconstructed IGL component for
configuration I (using combined evidence thresholding) and the middle
panel for configuration II (not using combined evidence thresholding). The
lower panel shows the surface brightness profiles of the input model
(continuous line) and two reconstructed IGL components (short dashed
and long dashed lines for the configurations I and II, respectively),
in arbitrary magnitude scale. A segment of the profiles is shown to
allow a better view of the differences.  It can be noticed that the
radial profile has a slightly sharper decrease for configuration II.
\label{figdistrec}}
\end{figure}

In reconstructing the IGL component alone (without galaxies) we find
that none of the cases I to IV presents a good description of the
{\it central} region of the simulated IGL exponential profile. In
other words, the central cuspy model is never reconstructed properly.
Indeed, the very central profile does not follow the original circular
light distribution and a radial distortion of the light profile can be
noticed, which is caused by the reconstruction process using the different
wavelet coefficients. This effect is stronger in the cases which do not
use the combined evidence thresholding.  The radial distortions for the
cases with and without the combined evidence thresholding are shown in
figure~\ref{figdistrec}. There are no differences from the reconstructions
for the cases with and without the restricted criterion.

In the third image configuration (galaxies+IGL), the galaxies
reconstruction presents the same characteristics of the case without the
IGL component, {\it i.e.}, they present a ``false'' common halo in all
four setups. There is a large difference in the reconstruction of the
central part of the IGL component. The light of this part is actually
included in the reconstruction of the galaxy component, to which it
does not represent any considerable increase in flux, because of the
low surface brightness of this component. The reconstruction of this
component suffers from all the same effects noticed before.

A second step in the analysis of the simulated images is to study the
radial profile of the IGL. Comparing the models with the reconstructed
objects, for the three different intensities of the IGL component,
with and without the galaxy component superposed, we can establish
to which light level or $S/N$ our results are reliable.  The central
regions were poorly reconstructed and we are not taking them into
account in this radial study. We have defined a cutoff radius for the
analysis, after which the reconstruction of the object is no longer
reliable. This radius was defined based on the flux differences between
the model and the reconstructed object in each radial ring analyzed. For
the maximum acceptable flux loss of the reconstructed object, we have
assumed a conservative limit of 10\%.  In each of the six images with
IGL component, this cutoff radius corresponds to a detected $S/N$ of
about 0.1. We have generated then a new set of images, with different
sky levels and sky noise levels, to test the significance of this limit.
With this new set of images we confirm the cutoff radius corresponding to
a $S/N$ limit of 0.1.  This way, we have defined a minimum detected $S/N$
of 0.1 as the detection limit for the reconstructed objects using the
OV\_WAV, which represents a great improvement over the previous studies
on detecting extended low surface brightness structures.  A $S/N$ of 0.1
may seem to be an unrealistically low value for the detected structures,
however we note that as the parts that compose the objects were detected
in the wavelet space with a 5-$\sigma$-detection level, these are indeed
significant, even though they are very faint in normal space.

We have analyzed the artificial images using the minimum $S/N$ cutoff
limit of 0.1 for the reconstructed objects and its corresponding models,
and the comparison presented consistent results, showing that the objects
reconstruction is very reliable within this limit of $S/N$. The results
can be seen in table~\ref{tabsimgloblim}.

\begin{table}

\centering
\caption{Simulation analysis results with cutoff in $S/N = 0.1$
\label{tabsimgloblim}}
\begin{tabular}{llrrrr}

\hline
      &       & \multicolumn{4}{c}{Parameters setup} \\
Image & Comp. & I & II & III & IV \\
\hline

Only Gal    & Gal     &  -0.1\% &   +0.1\% &    -2.3\% &     -0.2\% \\
\hline
Only IGL    & IGL A   &  -8.4\% &  -10.0\% &    -8.4\% &    -10.0\% \\
      & IGL B   &  -4.4\% &   -5.2\% &    -4.7\% &     -5.2\% \\
      & IGL C   &  -9.1\% &  -21.9\% &    -6.9\% &     -2.2\% \\
\hline
Gal+IGL   & Gal     &  +1.0\% &   +1.0\% &    +1.1\% &     +1.3\% \\
      & IGL A   & -11.1\% &  -12.7\% &   -11.9\% &    -13.7\% \\
      & Total A &  -2.0\% &   -2.4\% &    -2.0\% &     -2.4\% \\
      & Gal     &  +1.4\% &   +1.5\% &    +3.1\% &     +3.1\% \\
      & IGL B   & -14.1\% &  -12.1\% &   -26.4\% &    -25.6\% \\
      & Total B &  -0.6\% &   -0.2\% &    -0.6\% &     -0.5\% \\
      & Gal     &  +1.1\% &   +1.1\% &    +0.1\% &     +0.2\% \\
      & IGL C   & -37.9\% &  -27.5\% &   -33.5\% &    -50.9\% \\
      & Total C &  -1.0\% &   -0.4\% &    -1.7\% &     -2.5\% \\

\hline
\end{tabular}

Column (1) indicates the image type, described in the text.\\
Column (2) indicates the analyzed component: ``Gal'' for the galaxies
component; ``IGL'' for the diffuse intra-group light component, with three
different brightness A, B and C, from brighter to fainter; and ``Total''
for galaxies + IGL, with brightness A, B and C (see section 3.2.1b).\\
Columns (3-6) show the percentage of light difference from the
reconstructed to the model components for parameters setups I-IV, as
described in the text.  Positive signal indicates a light gain during
the reconstruction and negative signal indicates a light loss.

\end{table}

To estimate the errors, we have used a relation of the mean $S/N$ of the
reconstructed objects and its corresponding light loss.  For the galaxy
component, high surface brightness structures, the typical light loss
is of 1\%, and taking into consideration the possible contamination
effect discussed above, as a 2\% contamination, the galaxy component
final error is typically of 2.2\%.  To estimate the errors of the low
surface brightness structures, we have interpolated the $S/N$--light loss
relation obtained from our simulations.  A 10\% minimum error value was
defined to structures with mean $S/N$ higher than 0.5, not covered by the
simulations. At the faint-end, the simulated cases cover the detected
range of values.  For a complete description of the input parameters
of the OV\_WAV and the tests applied to define its confidence levels,
see \citet{epi05}.

\section{Analysis and Results}

Based on the simulations results, we chose to use the restricted criterion
and the combined evidence thresholding (case I), which gave the best IGL
reconstruction results.  The procedure for each group is described below.

\subsection{HCG 79}

\begin{figure*}
\centering
\includegraphics[width=8.0cm]{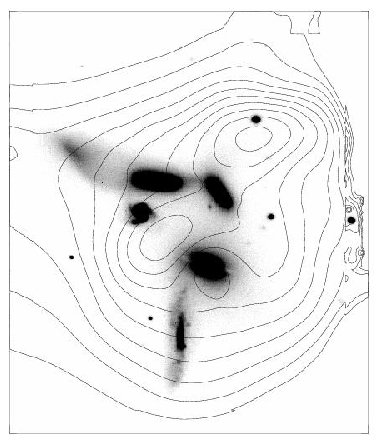}
\includegraphics[width=8.0cm]{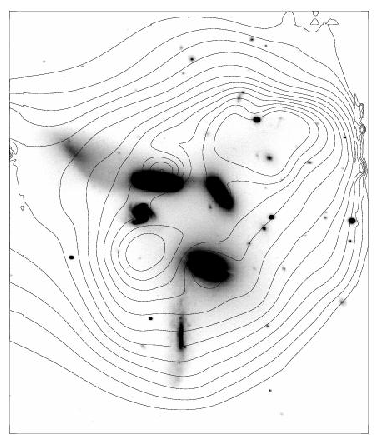}
\caption{IGL component of HCG 79 identified and reconstructed with the
OV\_WAV package. The left panel shows the image in the $B$ band and the
IGL component as contour curves with surface brightness levels which range
from $24.2$ to $25.1$ magnitudes in steps of $0.1~B~mag~arcsec^{-2}$.
The right panel shows the image in the $R$ band and the IGL component as
contour curves with surface brightness levels which range from $23.2$
to $24.3$ magnitudes in steps of $0.1~R~mag~arcsec^{-2}$.
\label{figdifh79}}
\end{figure*}

The HCG 79 images in the $B$ and $R$ bands were deconvolved in 11 wavelet
coefficients ($2^{10}$ pixels, about the size of the image).  As described
before, the objects were reconstructed in multiple iterations.  Using the
results of the multiple iterations we can recompose, separately, the group
galaxies and faint structures which form the IGL component of this group,
as can be seen in figure~\ref{figdifh79}.

We have detected a irregular shape IGL component, that represents
$46\pm11$\% of the total light in the $B$ band and $33\pm11$\% in the $R$
band, which corresponds to an apparent magnitude of $B = 14.0\pm0.16$
and $R = 13.1\pm0.15$, up to the surface brightness detection limits of
$\mu_B = 28.4$ and $\mu_R = 27.8$. This is a bright IGL component with
a mean $S/N$ of $2.8$ in the $B$ band and $3.6$ in the $R$ band, much
brighter than the simulated cases.  The detection limits correspond to the
surface brightness of $0.1\cdot \sigma_{Sky}$ ($S/N = 0.1$) in each band.

The mean surface brightness of the IGL component is $\mu_B = 24.8\pm0.16$
and $\mu_R = 23.9\pm0.15$, which is $2.6$ magnitudes in $B$ and
$2.9$ magnitudes in $R$ fainter than the sky brightness, while the
detection limits are $6.2$ and $6.7$ magnitudes fainter than the sky
brightness levels in $B$ and $R$, respectively. The IGL component has
a mean color $(B-R)_0 = 0.86\pm0.22$, which is bluer than the galaxies'
component that has $(B-R)_0 = 1.47\pm0.15$ (typical color for early-type
galaxies), and bluer than the expected value for an intracluster component
\citep{tre98,zib05,som05} formed by an old population with no significant
on-going star formation, but not very different from the blue sources
detected by \citet{ada05} in the Coma cluster. The extinction corrections
were made using \citet{rie85} extinction laws and \citet*{sch98}
extinction maps.  Measurements of the IGL and galaxy components given
above were performed in the same areas for the $B$ and $R$ bands.

Taking into account the IGL component in each of the groups we
recalculated the mass to light ratios ($M/L$) for HCG 79 and
HCG 95, correcting the values estimated by \citet{hic92} for the
previously undetected luminous component.  We note, however, that these
determinations of mass-to-light ratio are very uncertain, given that the
spectroscopic mass of the group was obtained from velocities of only a
few group members. No X-ray mass estimates were found for these groups
in the literature.  For HCG 79 the $M/L$ drops from $11.3$ to $6.2$.

\subsection{HCG 95}

\begin{figure*}
\centering
\includegraphics[width=8.0cm]{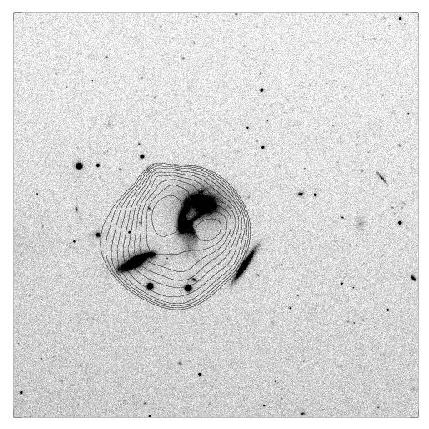}
\includegraphics[width=8.0cm]{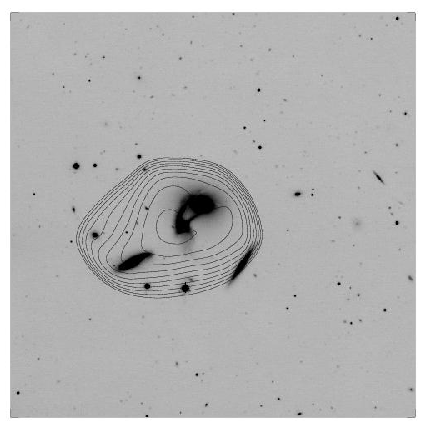}
\caption{IGL component of HCG 95 identified and reconstructed with the
OV\_WAV package. The left panel shows the image in the $B$ band and the
IGL component as contour curves with surface brightness levels which range
from $26.9$ to and $27.8$ magnitudes in steps of $0.1~B~mag~arcsec^{-2}$.
The right panel shows the image in the $R$ band and the IGL component
as contour curves with surface brightness levels which range from $24.9$
to $26.7$ magnitudes in steps of $0.2~R~mag~arcsec^{-2}$.
\label{figdifh95}}
\end{figure*}

The same procedure was applied to this group and the results can be seen
in figure~\ref{figdifh95}.

In the case of HCG 95, the results were analyzed disconsidering the
discordant-redshift galaxy HCG 95B, which is not part of the group,
as mentioned in Section 2.1.3. We have detected an almost spherical IGL
component, much less prominent than the one in HCG 79, concentrated around
the system HCG 95A/C. This is not surprising, given that there are strong
interactions going on in this group - in fact the interaction between A
and C may be the main source of this IGL component. We estimated that
$11\pm26$\% of the total light in the $B$ band and $12\pm10$\% in the
$R$ band are in this IGL component, which corresponds to an apparent
magnitude of $B = 16.9\pm0.30$ and $R = 15.1\pm0.15$, up to the limiting
surface brightness of $\mu_B = 28.2$ and $\mu_R = 28.1$. In this case,
for the $B$ band, the mean $S/N$ is only $0.2$, and for the $R$ band,
a brighter IGL component was detected with mean $S/N$ of $1.2$.

The IGL component has a mean surface brightness of $\mu_B = 27.3\pm0.30$
and $\mu_R = 25.5\pm0.15$, which is 5.0 and 5.1 magnitudes fainter that
the sky levels for the $B$ and $R$ bands, respectively. The detection
limits are 5.9 magnitudes fainter than the sky level in the $B$ band and
7.8 magnitudes in the $R$ band. The mean color of the IGL is $(B-R)_0
= 1.75\pm0.34$ and the galaxies have $(B-R)_0 = 1.61\pm0.15$. In this
group, both the galaxies and the IGL have colors consistent with those
of old stellar populations.  The tidal tails of the system HCG 95A/C
were considered as part of the galaxy component not as part of the IGL.

Also for this group we recalculate the $M/L$, taking into account the
intra-group component, which drops from $35.1$ \citep{hic92} to $31.2$.

\subsection{HCG 88}

\begin{figure*}
\centering
\includegraphics[width=8.0cm]{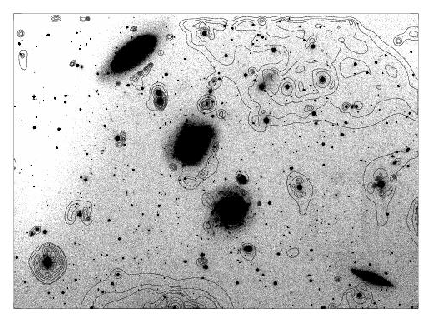}
\includegraphics[width=8.0cm]{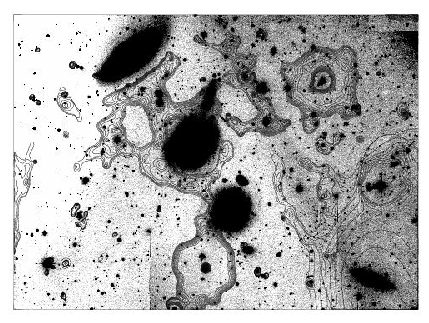}
\caption{Residual identifications in the HCG 88 images with the OV\_WAV
package. The left panel shows the image in the $B$ band and the sky
as contour curves with surface brightness levels ranging from $24.4$
to $26.2$ magnitudes in steps of $0.2~B~mag~arcsec^{-2}$.  The right
panel shows the image in the $R$ band and the sky as contour curves with
surface brightness levels ranging from $24.4$ to $26.2$ magnitudes in
steps of $0.2~R~mag~arcsec^{-2}$.
\label{figdifh88}}
\end{figure*}

The same procedure above was also applied to this group, but in this case,
no IGL component was detected up to a surface brightness detection limit
of $\mu_B = 29.1$ and $\mu_R = 27.9$. This is not unexpected, given
that the group is composed by four late-type galaxies which contain
90\% of their H{\sc i} component still attached to the galaxies disks
\citep{ver01}. Figure~\ref{figdifh88} shows the group and the detected
sky level, showing that no IGL is present in this group.

Table~\ref{tabres} summarizes the properties of the detected IGL
components in our sample.

\begin{table*}

\centering
\caption{Properties of the IGL component detected in our sample.
\label{tabres}}
\begin{tabular}{llllllll}

\hline
Group & \multicolumn{2}{c}{\% ($B$ and $R$)} & \multicolumn{2}{c}{$\mu$ ($B$ and $R$)} & \multicolumn{2}{c}{Mag. ($B$ and $R$)} & $(B-R)_0$ \\
\hline

HCG 79  & $46\pm11$\% & $33\pm11$\% & $24.8\pm0.16$ & $23.9\pm0.16$ & $14.0\pm0.16$ & $13.1\pm0.16$ & $0.86\pm0.22$ \\
HCG 88  & ND         & ND         & ND            & ND            & ND            & ND            & ND            \\
HCG 95  & $11\pm26$\% & $12\pm10$\% & $27.3\pm0.30$ & $25.5\pm0.15$ & $16.9\pm0.30$ & $15.1\pm0.15$ & $1.75\pm0.34$ \\
\hline
\end{tabular}

Column (1) Group studied.\\
Columns (2-3) Fraction of the group's total light in the IGL component, bands 
$B$ and $R$, respectively.\\
Columns (4-5) Mean surface brightness of the IGL component, bands $B$ and $R$, 
respectively.\\
Columns (6-7) Integrated magnitude of the IGL component, bands $B$ and $R$, 
respectively.\\
Column (8) Extinction corrected color of the IGL component.
\end{table*}

\section{Discussion and Conclusions}

We present here a short summary of the results we have obtained, before
we proceed with the discussion:

(1) A new approach was presented for the detection of extended low
surface-brightness structures, such as the IGL component in compact
groups of galaxies, through wavelet analysis, using the OV\_WAV package.

(2) We have detected a prominent irregular IGL component around HCG 79,
which represents $46\pm11$\% and $33\pm11$\% of the total light of the
group in the B and R bands, respectively. The mean color of the IGL
in this group is $(B-R)_0 = 0.86\pm0.22$, significantly bluer than the
color of the galaxies in the group.

(3) HCG 95 presents a fainter IGL component, representing $11\pm26$\%
and $12\pm10$\% of the group's total light in the $B$ and $R$ bands,
respectively, with a mean color $(B-R)_0 = 1.75\pm0.34$.

(4) The filamentary group HCG 88 presented no IGL component, down to the
surface brightness detection limit, as expected for a group composed by
late-type galaxies with most of its H{\sc i} still in the galaxies disks.

Using the OV\_WAV package we were able to detect the IGL component of
compact groups of galaxies, a very difficult task to be performed with the
usual algorithms, for which it is necessary to specify the sky brightness
levels and the stars/galaxies models. The usual way to determine the IGL
component, performed by \citet{nis00} for HCG 79, for an example, is to
model the galaxies with {\tt STSDAS.ELLIPSE} and subtract them from the
image (with {\tt STSDAS.BMODEL}). Then one just measures the remaining
light above a chosen sky level, and assign it to the IGL component.
Another method used, in this case for HCG 90, was that by \citet{whi03},
of isophotal analysis: after sky subtraction, all the light below a
certain surface brightness level was considered as IGL.  Those technics
are highly undesirable specially in cases such as those of HCG 79 or HCG
95, where elliptical models (fit by {\tt STSDAS.ELLIPSE}) are a very bad
representation for the morphology of the galaxies, and the presence of
bright interacting features may increase the isophotal analysis ambiguity.
With the OV\_WAV, the IGL detection could be performed without any ``a
priori'' information about the shapes of the galaxies or the sky level.

An evolutionary sequence can be envisioned, for our small sub-sample
of compact groups, where HCG 79 is the more evolved group, HCG 95 is in
an intermediate stage of evolution and HCG 88 is a group that has just
started its evolution as a group. We give details of each group below.

\subsection {HCG 79}

The IGL component detected in HCG 79, a group which show several
indications of been ``about to collapse'' in a single structure, or at
an advanced evolutionary stage, has a light envelope with irregular form
which takes up about one third to half the total group light. Such a
high contribution from the IGL component is in contrast with the value
of 13\% found by \citet{nis00}, for the same group, from independent
$VR$ and $I$-band data, although the morphology of the IGL was found
to be very similar.  The disagreement probably comes from the technic
differences described above.

We can notice in figure~\ref{figdifh79} a strong light concentration to
the northwest, opposite to the tidal debris located to the northeast,
which could have being generated in the same interaction process, like
symmetrical tidal tails, but with a smaller amount of mass.

The irregular shape of the IGL may be an indication that the group
potential is not relaxed in an spherical shape. On the other hand, the
amount of light in the IGL component of HCG 79 suggests that the group
has probably already been in a compact configuration for some time.

The X-rays halo observed with ROSAT by \citet{pil95b} (in a 3$\sigma$
level), is spatially coincident with the IGL component, indicating that
this is following the gravitational potential and it is tracing the dark
matter halo of the group. The H{\sc i} of the group is concentrated in
two clouds, one in the southern part, below the only late-type galaxy
(HCG 79D), and one in the region of the northeastern tidal debris.
These were probably a result of an interaction of this late-type galaxy
with the group halo \citep{wil91,ver01}.

According to \citet{pal02} this group presents on one hand signs of high
dynamical evolution, specially in the optical, since tidal structures
in broad band optical images are clearly seen, and signs of almost no
evolution, on the H{\sc i} content, which is basically associated to
the disk of the only late-type galaxy (HCG 79D), which leads to the
absence of interaction triggered star formation. The star clusters
found in \citet{pal02} are red and even though they may have been
formed during interaction processes, they are old clusters ($10^{8.5}$
to $10^{9.5}$ yr). No tidal dwarf galaxies and just a few dwarfs were
detected. \citet{pal02} classified this group as being at an intermediate
stage of evolution (they used the expression ``beginning of the end''
to describe the evolutionary stage of this group). This classification
also agrees with the findings of \citet*{coz04}, who used a relation
between group velocity dispersion and galaxy activity.  An IGL component
of this group is mentioned as an optical evidence of dynamical evolution,
and our high fraction of IGL indicates that this group is really half
way to collapse or a step further.

The mean color of the galaxies in HCG 79 is consistent with the expected
old stellar population of a group dominated by early-type galaxies,
while the IGL component is significantly bluer.  The blue color of the
IGL in HCG 79 is inconsistent with the observations and simulations
for galaxy clusters \citep{tre98,som05} and even for higher redshifts
studies as \citet{zib05} at $z \approx 0.25$ in the SDSS clusters.
The stellar populations which constitute the diffuse light could have
been partially stripped from the outer parts of the group galaxies,
during interaction processes, as the bright tidal debris described before
and partially due to the destruction of dwarf galaxies with bluer colors
than the giant group members. The destruction of dwarf galaxies by the
group tidal field is one of the possible mechanisms of formation of the
bimodal globular cluster population usually found in early-type galaxies
and may contribute to the IGL component as well \citep*{hil99}. No blue
dwarf galaxy was detected in this group by \citet{pal02}. Blue diffuse
regions were detected by \citet{ada05} in the Coma cluster, where spiral
galaxies were disrupted or harassed, leading to infall of star-forming
material to the cluster potential.

The fraction of light in the IGL component of HCG 79 is $M_B =
-20.2\pm0.6$, at the groups distance, which corresponds to a galaxy
twice as large as HCG 79A, the brightest group galaxy. Tidal stripping
of the galaxies' outer parts is the most probable explanation for the
small size of those galaxies, about a third of the size of a normal disk
galaxy \citep{wil91}.

\subsection {HCG 95}

HCG 95 presents a nearly spherical IGL component centered on HCG 95A.
This group shows several tidal features indicating clear interaction and
has an IGL component which corresponds to about 10\% of the total light
of the group. This is most probably formed by the stripping of material
from the interacting galaxies.

This group was detected in X-rays by ROSAT \citep{pon96} with a flux
upper limit of $log L_X = 42.14~erg~s^{-1}$, indicating a shallow
potential well, typical of spiral-dominated groups and was also detected
in infrared by IRAS, with emission centered on HCG 95C and enclosing the
whole group \citep{all96}. HCG 95C1 and C2, HCG 95D and two previously
unknown dwarf galaxies were detected in H{\sc i}. In addition, radio
continuum emission was searched for within a field of 9 arcminutes
around the group and HCG 95C (centered on C1) was detected as a continuum
source \citep{huc00,ver01}.

HCG 95 presents clear signs of interactions as tidal tails and bridges,
${\rm H{\alpha}}$ emission and TDG candidates on the main tidal tail
\citep{igl01}. The interaction system (HCG 95A/C), which is indeed
formed by three galaxies (an elliptical, HCG 95A, and two disk galaxies,
HCG 95C1 and C2) giving rise to the complex arrangement of tidal tails
observed in this group, was studied by \citet{igl97,igl98} and two models
were proposed.  The first one suggests that the two disk galaxies (C1
and C2) are in on-going interaction and this pair interacts with the
elliptical galaxy. The second scenario puts forward that the two disk
galaxies interact independently with the elliptical galaxy in different
times, generating a pair of tails each and are projected over each other
\citep[see][ for a more complete description of the models]{igl97,igl98}.

The on-going interactions suggest that this group is young, or entering an
intermediate stage of dynamical evolution. \citet{igl97}, found unusual
color and patches of dust in the elliptical galaxy, which together with
the ${\rm H{\alpha}}$ emission would be a sign of mass transfer among the
galaxies through the tidal tails. The authors argue that those effects
where indicating that this interacting system would probably merge in
a few orbital periods.

The detected IGL component has a magnitude of $M_B = -19.4\pm0.6$,
which is about one third of the total luminosity of HCG 95A. This result
gives support to the scenario that puts forward that this group is in
an intermediate evolutionary stage, since the IGL component has most
probably been created by the interaction of HCG 95A/C. No extension of
the IGL was detected towards the discordant galaxy HCG 95B, an additional
indication that this galaxy does not belong to the group.

\subsection {HCG 88}

No IGL component was detected for HCG 88 down to the surface brightness
detection limit of $\mu_B = 29.1$ and $\mu_R = 27.9$.  This result
supports the scenario of HCG 88 being a filamentary system which has just
initiated its evolution. Further support to this scenario comes from the
H{\sc i} observations for the group galaxies: all four late-type galaxies
of the group still contain 90\% of the H{\sc i} attached to their disks
\citep{ver01} suggesting this is a group that has been together for a
short time.

The very low velocity dispersion of this group, $31~km~s^{-1}$, lower than
the expected for a virialized structure with the group characteristics,
was used by \citet{mam00} to argue that this group would be a chance
alignment of galaxies in a turnaround point. This group was also pointed
out as a good candidate for \citet*{her95} projected filamentary groups.

On the other hand, according to \citet{ver01}, a group with a filamentary
structure such as HCG 88, with very low velocity dispersion and
short crossing time \citep[very different from the predicted from][
simulations]{her95}, with a high fraction of H{\sc i} associated to
the galaxy disks and very isolated \citep[there are only two galaxies,
about 3 magnitudes fainter than the faintest giant member galaxy, HCG
88D, in a 700 kpc-side square centered on the group,][]{dec97} is very
improbable to be a chance alignment.  Our conclusion that the group is in
an initial evolutionary stage agrees with the results from \citet{pla03}
and \citet{coz04}.  Dynamical friction could also be responsible for
low observed group velocity dispersion, but it would only be expected
in a group in advanced stage of collapse, which does not seem to be the
case here.

\subsection{Final Remarks}

No significant relations between the structure size and the intra-group or
intracluster light fraction were found. The literature points to values
ranging from 16 to 28\% for clusters with richness 2 \citep{fel04},
13\% for a cluster with richness 3 \citep{fel02}, an average of 11\% for
clusters at $z\approx0.25$ \citep{zib05}, 17 to 43\% in Virgo, depending
on the studied region \citep{arn02} and 50\% in Coma \citep{ber95}. The
fraction of IGL found for HCG 79 is a high value. It is interesting
to note that the values for clusters may include the light of several
galaxies which are not in reality involved in the interaction processes
that generate the intracluster component, but in compact groups,
in particular for HCG 79, all the member galaxies seem to be somehow
involved in the interactions.

We could recalculate the $M/L$ for HCG 79 and HCG 95, estimated by
\citet{hic92}, correcting the luminous component by the IGL detected
in this work. However, these measurements are very uncertain given
that the only mass determinations available for these groups are from
the spectroscopy of a few member galaxies \citep{hic92}.  Therefore,
any attempt to find correlations between mass, or $M/L$, and the IGL
fraction would be speculative at this point.

\citet{hic92} point out a possible relation where shorter group
crossing times would indicate a more evolved stage, estimated by the
spiral fraction and the magnitude difference between the first and the
second ranked galaxies ($\Delta m_{12}$). Taking the IGL fraction as
a dynamical evolution indicator, the groups studied here do follow
the same trend, however without any statistical significance given
the size of our sample (three groups).  HCG 79 has a crossing time of
$0.004~H_{0}^{-1}$ and an IGL fraction of $46\pm11$\%, HCG 95 has a
crossing time of $0.007~H_{0}^{-1}$ and an IGL fraction of $11\pm26$\%
and HCG 88 has a crossing time of $8.7~H_{0}^{-1}$ and no IGL detection.

The study of the relations between the IGL characteristics and the group
environment will be addressed in a forthcoming paper when a statistically
significant sample can be studied.

In conclusion, with the presently studied sample we could define an
evolutionary sequence, in which HCG 79 is the more evolved, in an
about-to-collapse stage; HCG 95 is an intermediate to young group in
this sequence; and finally HCG 88 is the least evolved of the groups,
which has just started its dynamical evolution.

\section{Acknowledgments}

We would like to thank Mike Bolte for obtaining the images, doing the
pre-reduction and the photometric calibration, Hugo Capelato and Bodo
Ziegler for the long science discussions and support to this project,
Carlos Raba\c ca and Daniel Epit\'acio Pereira for the software
development and support and Francisca Brasileiro for taking the
spectrum of HCG 95B. We also would like to thank the referee of this
paper, Dr. Christophe Adami, for useful comments which improved this
manuscript considerably. CDR is supported by FAPESP (Funda\c c\~ao de
Amparo a Pesquisa do Estado de S\~ao Paulo) grants No. 96/08986--5 and
02/06881--4 and CAPES/DAAD (Coordena\c c\~ao de Aperfei\c coamento de
Pessoal de N\'{\i}vel Superior/Deutscher Akademischer Austausch Dienst)
grant No. BEX: 1380/04--4.  CMdO would like to acknowledge support
from the Brazilian agencies FAPESP (projeto tem\'atico 01/07342-7),
CNPq and CAPES.

\label{lastpage}

\end{document}